\documentclass[11pt]{article}
\usepackage{verbatim,color,amssymb,epsfig}
\usepackage[usenames,dvipsnames]{xcolor}
\usepackage{fancyhdr}
\usepackage[authoryear,sort]{natbib}
\usepackage{latexsym,epsfig,amssymb,amsmath,amsfonts,graphicx,amsthm,ifthen,pifont,comment,enumerate,setspace,multirow,color}
\usepackage{bm}
\usepackage[ruled,vlined]{algorithm2e}
\usepackage{epstopdf}
\usepackage{float}
\usepackage{stackengine}
\usepackage{xcolor}
\usepackage{tikz}
\theoremstyle{definition}
\newtheorem{definition}{Definition}[section]
\usepackage{nameref,hyperref}
\usepackage{longtable}
\usepackage{framed}
\usetikzlibrary{bayesnet}
\usepackage{multicol,booktabs,colortbl,tabularx}
\hypersetup{
	colorlinks=true,       % false: boxed links; true: colored links
	linkcolor=blue,        % color of internal links
	citecolor=blue, 	     						  % color of links to bibliography
	filecolor=magenta,     % color of file links
	urlcolor=blue         
}
\begin{document}
\thispagestyle{empty}
\baselineskip=28pt
{\LARGE{ \begin{center}
     \textbf{Estimation of COVID-19 spread curves integrating global data and borrowing information} 
\end{center}}}

\baselineskip=12pt

\vskip 10mm
\begin{center}
\begin{large}
By SE YOON LEE, BOWEN LEI, and BANI K. MALLICK
\end{large}
\\
$\quad$
\\
\emph{Department of Statistics, Texas A\&M University, College Station, Texas, 77843, U.S.A.}
\\
seyoonlee@stat.tamu.edu$\quad$bowenlei@stat.tamu.edu$\quad$bmallick@stat.tamu.edu
\end{center}

\vskip 10mm
%\begin{center}
\begin{center}
    \Large{\bf Abstract} 
\end{center}
{\noindent 
Currently, novel coronavirus disease 2019 (COVID-19) is a big threat to global health. The rapid spread of the virus has created pandemic, and countries all over the world are struggling with a surge in COVID-19 infected cases. There are no drugs or other therapeutics approved by the US Food and Drug Administration to prevent or treat COVID-19: information on the disease is very limited and scattered even if it exists. This motivates the use of data integration, combining data from diverse sources and eliciting useful information with a unified view of them. In this paper, we propose a Bayesian hierarchical model that integrates global data for real-time prediction of infection trajectory for multiple countries. Because the proposed model takes advantage of borrowing information across multiple countries, it outperforms an existing individual country-based model. As fully Bayesian way has been adopted, the model provides a powerful predictive tool endowed with uncertainty quantification. Additionally, a joint variable selection technique has been integrated into the proposed modeling scheme, which aimed to identify possible country-level risk factors for severe disease due to COVID-19.
%\end{center}
}
\baselineskip=12pt

\baselineskip=12pt
\par\vfill\noindent

\par\medskip\noindent
\clearpage\pagebreak\newpage
\pagenumbering{arabic}
\newlength{\gnat}
\setlength{\gnat}{22pt}
\baselineskip=\gnat

% Use "Eq" instead of "Equation" for equation citations.
\section{Introduction}\label{sec:Introduction}
\begin{figure}[h]
    \centering
    \includegraphics[scale = 0.45]{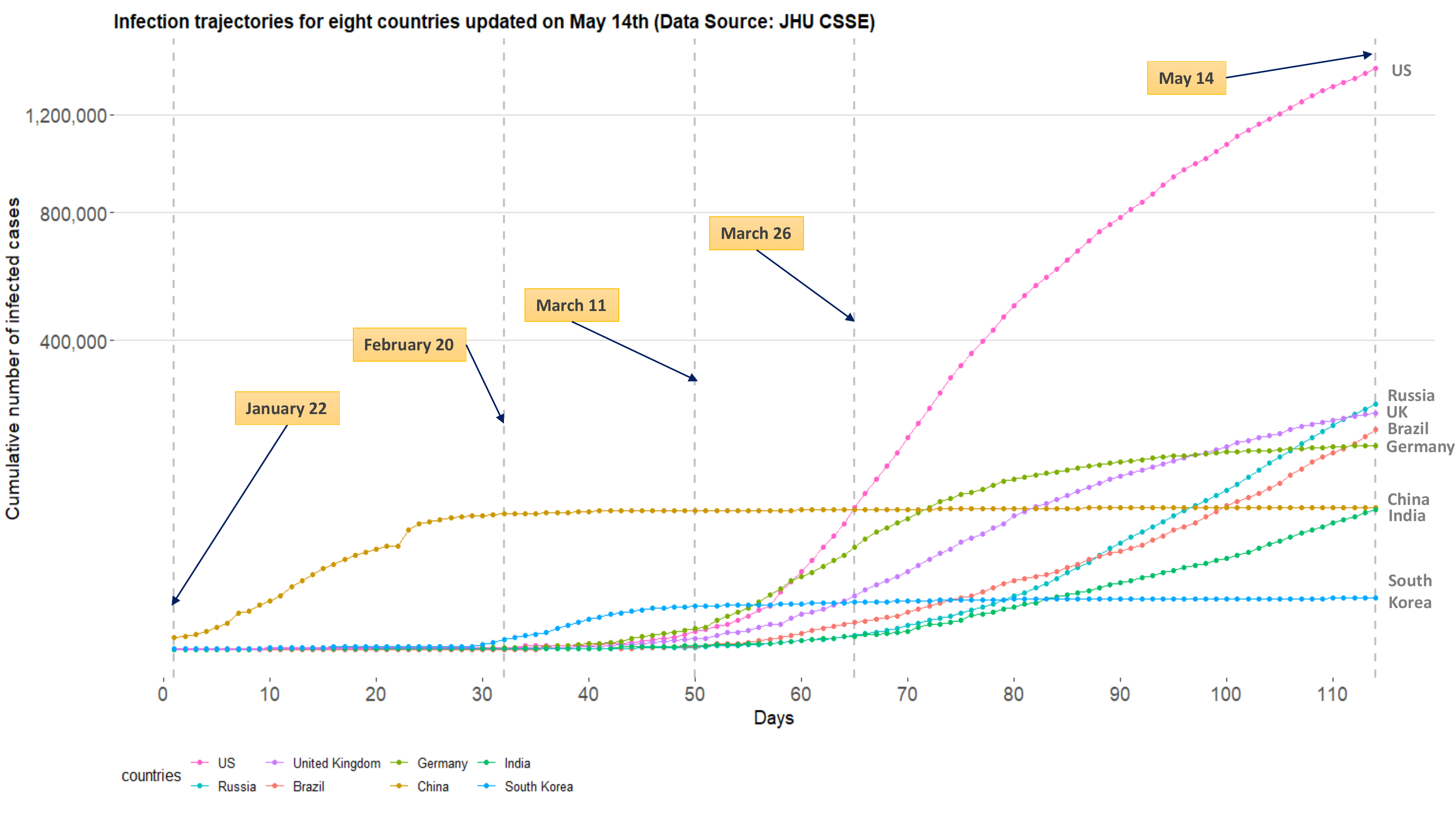}
    \caption{\baselineskip=10pt  Daily trajectories for cumulative numbers of COVID-19 infections for eight countries (US, Russia, UK, Brazil, Germany, China, India, and South Korea) from January 22nd to May 14th. (Data source: Johns Hopkins University CSSE)}
    \label{fig:6_countries}
\end{figure}

\begin{figure}[h]
    \centering
    \includegraphics[scale = 0.45]{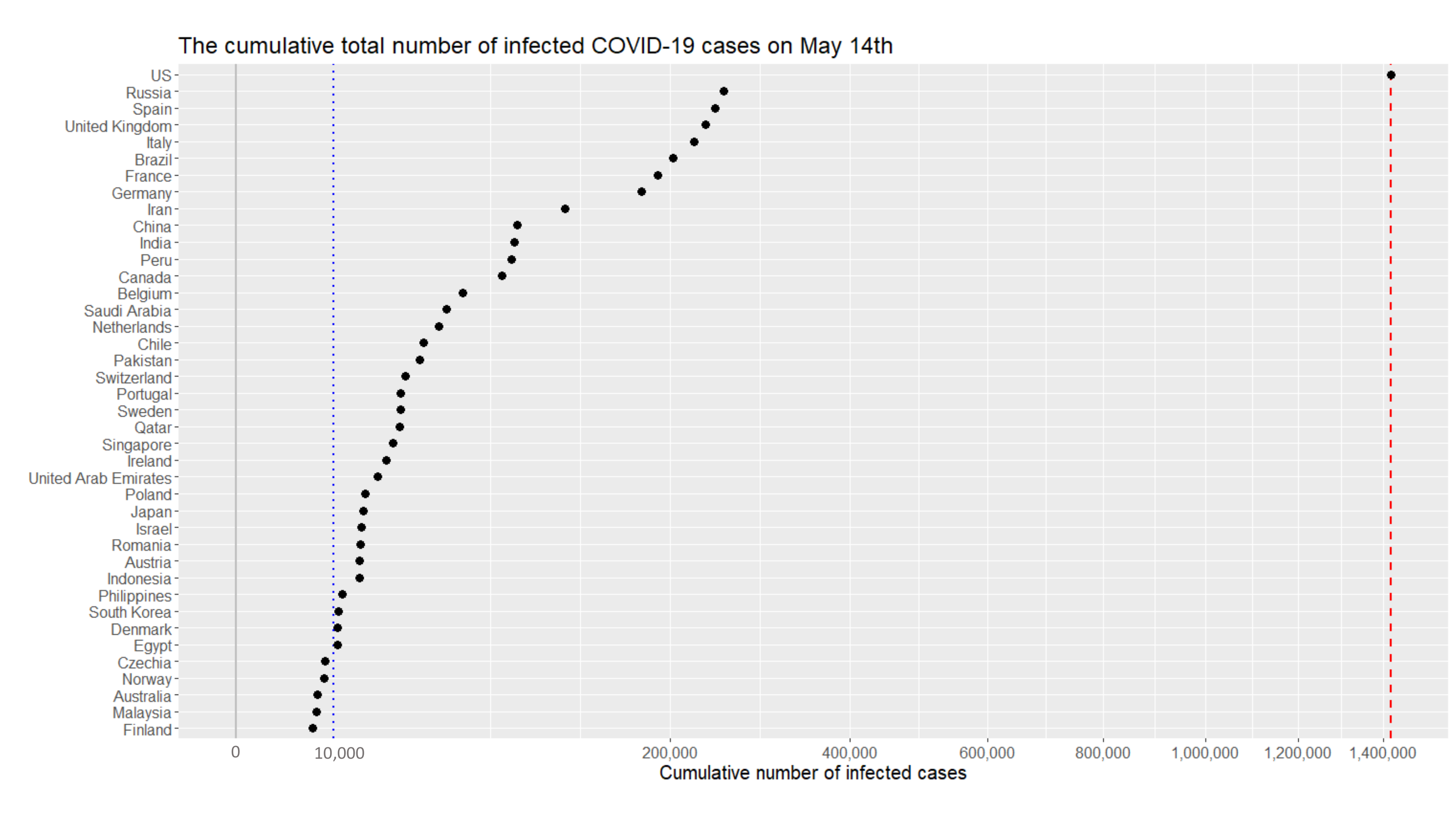}
    \caption{\baselineskip=10pt  Cumulative numbers of infected cases for 40 countries on May 14th. ($x$-axis are scaled with squared root for visualization purpose.) The red dashed vertical lines represents 1,417,774 cases.}
    \label{fig:Confirmed cases on May 14th}
\end{figure}

Since Thursday, March 26, 2020, the US leads the world in terms of the cumulative number of infected cases for a novel coronavirus, COVID-19. On this day, a dashboard provided by the Center for Systems Science and Engineering (CSSE) at the Johns Hopkins University (\href{https://systems.jhu.edu/research/public-health/ncov/}{https://systems.jhu.edu/-}) \citep{dong2020interactive} reported that the numbers of the confirmed, death, and recovered from the virus in the US are 83,836, 1,209, and 681, respectively. Figure \ref{fig:6_countries} displays daily infection trajectories describing the cumulative numbers of infected cases for eight countries (US, Russia, UK, Brazil, Germany, China, India, and South Korea), spanning from January 22nd to May 14th, which accounts for 114 days. The dotted vertical lines on the panel mark certain historical dates that will be explained. As seen from the panel, the US has been a late-runner until March 11th in terms of the infected cases, but the growth rate of the cases had suddenly skyrocketed since the day, and eventually excelled the forerunner, China, just in two weeks, on March 26th. Figure \ref{fig:Confirmed cases on May 14th} shows the cumulative infected cases for 40 countries on May 14th: on the day, the number of cumulative infected cases for the US was 1,417,774 which is more than five times of that of Russia, 252,245. 

%%%%%%%%%%%%%%%%%%%%%%%%%%%%%%%%%

%%%%%%%%%%%%%%%%%%%%%%%%%%%%%%%%%

%%%%%%%%%%%%%%%%%%%%%%%%%%%%%%%%%

%%%%%%%%%%%%%%%%%%%%%%%%%%%%%%%%%

Since the COVID-19 outbreak, there have been numerous research works to better understand the pandemic in different aspects \citep{EstimateDiamondship,2002.06563,JTD36385,2002.12298,REMUZZI2020,2003.05447,trendforecastchina,gao2020breakthrough}. Some of the recent works from statistics community are as follows. \cite{EstimateDiamondship} focused on a serial interval (the time between successive cases in a chain of transmissions) and used the gamma distribution to study the transmission on Diamond Princess cruise ship. \cite{2002.06563} proposed the generalized susceptible exposed infectious removed model to predict the inflection point for the growth curve, while \cite{JTD36385} modified the proposed model and considered the public health interventions in predicting the trend of COVID-19 in China. \cite{2002.12298} proposed a differential equation prediction model to identify the influence of public policies on the number of patients. \cite{trendforecastchina} used a symmetrical function and a long tail asymmetric function to analyze the daily infections and deaths in Hubei and other places in China. \cite{REMUZZI2020} used an exponential model to study the number of infected patients and patients who need intensive care in Italy. One of the major limitations of these works is that the researches are confined by analyzing data from a single country, thereby neglecting the global nature of the pandemic.

One of the major challenges in estimating or predicting an infection trajectory is the heterogeneity of the country populations. It is known that there are four stages of a pandemic: visit \href{https://economictimes.indiatimes.com/news/politics-and-nation/coronavirus-the-four-stages-of-a-global-pandemic/pandemic/slideshow/74884293.cms}{economictimes.indiatimes.com/-}. The first stage of the pandemic contains data from people with travel history to an already affected country. In stage two, we start to see data from local transmission, people who have brought the virus into the country transmit it to other people. In the third stage, the source of the infection is untraceable. In stage four the spread is practically uncontrollable. In most of the current literature, estimation or prediction of the infection trajectory is based on a single country data where the status of the country falls into one of these four stages. Hence, such estimation or prediction may fail to capture some crucial changes in the shape of the infection trajectory due to a lack of knowledge about the other stages. This motivates the use of data integration \citep{lenzerini2002data,huttenhower2006bayesian} which combines data from different countries and elicits a solution with a unified view of them. This will be particularly useful in the current context of the COVID-19 outbreak. 

Recently, there are serious discussions all over the world to answer the crucial question:  ``even though the current pandemic takes place globally due to the same virus, why infection trajectories of different countries are so diverse?" For example, as seen from Figure \ref{fig:6_countries}, the US, Italy, and Spain have accumulated infected cases within a short period of time, while China took a much longer time since the onset of the COVID-19 pandemic, leading to different shapes of infection trajectories. It will be interesting to find a common structure in these infection trajectories for multiple countries, and to see how these trajectories are changing around this common structure. Finally, it is significant to identify the major countrywide covariates which make infection trajectories of the countries behave differently in terms of the spread of the disease.

\section{Methods}\label{sec:Methods}
\subsection{Richards growth curve models}\label{subsec:Richards growth curve}
Richards growth curve model \citep{richards1959flexible}, so-called the generalized logistic curve \citep{nelder1962182}, is a growth curve model for population studies in situations where growth is not symmetrical about the point of inflection \citep{seber2003nonlinear,anton1988calculus}. The curve was widely used to describe various biological processes \citep{werker1997modelling}, but recently adapted in epidemiology for real-time prediction of outbreak of diseases; examples include SARS \citep{hsieh2004sars,hsieh2009richards}, dengue fever \citep{hsieh2009intervention,hsieh2009turning}, pandemic influenza H1N1 \citep{hsieh2010pandemic}, and COVID-19 outbreak \citep{wu2020generalized}.

There are variant reparamerized forms of the Richards curve in the literature \citep{causton1969computer,birch1999new,kahm2010grofit,cao2019new}, and we shall use the following form in this research
\begin{align}
\label{eq:Richards growth curve}
f(t ; \theta_1, \theta_2, \theta_3, \xi) &= \theta_1 \cdot [ 1 + \xi \cdot \exp \{-\theta_2 \cdot ( t - \theta_3) \}   ]^{-1/\xi},
\end{align}

where $\theta_1$, $\theta_2$, and $\theta_3$ are real numbers, and $\xi$ is a positive real number. The utility of the Richards curve (\ref{eq:Richards growth curve}) is its ability to describe a variety of growing processes, endowed with strong flexibility due to the shape parameter $\xi$ \citep{birch1999new}: analytically, the Richards curve (\ref{eq:Richards growth curve}) (i) becomes the logistic growth curve \citep{tsoularis2002analysis} when $\xi=1$, and (ii) converges to  Gompertz growth curve \citep{gompertz1825xxiv} as the $\xi$ converges to zero from positive side of real numbers. (Gompertz curve is $g(t ; \theta_1, \theta_2, \theta_3) = \theta_1 \cdot \exp\ [-\exp\  \{-\theta_2 \cdot (t - \theta_3)\}]$.) But it is also known that estimation of $\xi$ is a complicated problem \citep{wang2016study}, and we resort to a modern sampling scheme, elliptical slice sampler \citep{murray2010elliptical}, to estimate the $\xi$. (See SI Appendix for more detail.)

\begin{figure}[h]
    \centering
    \includegraphics[scale=0.43]{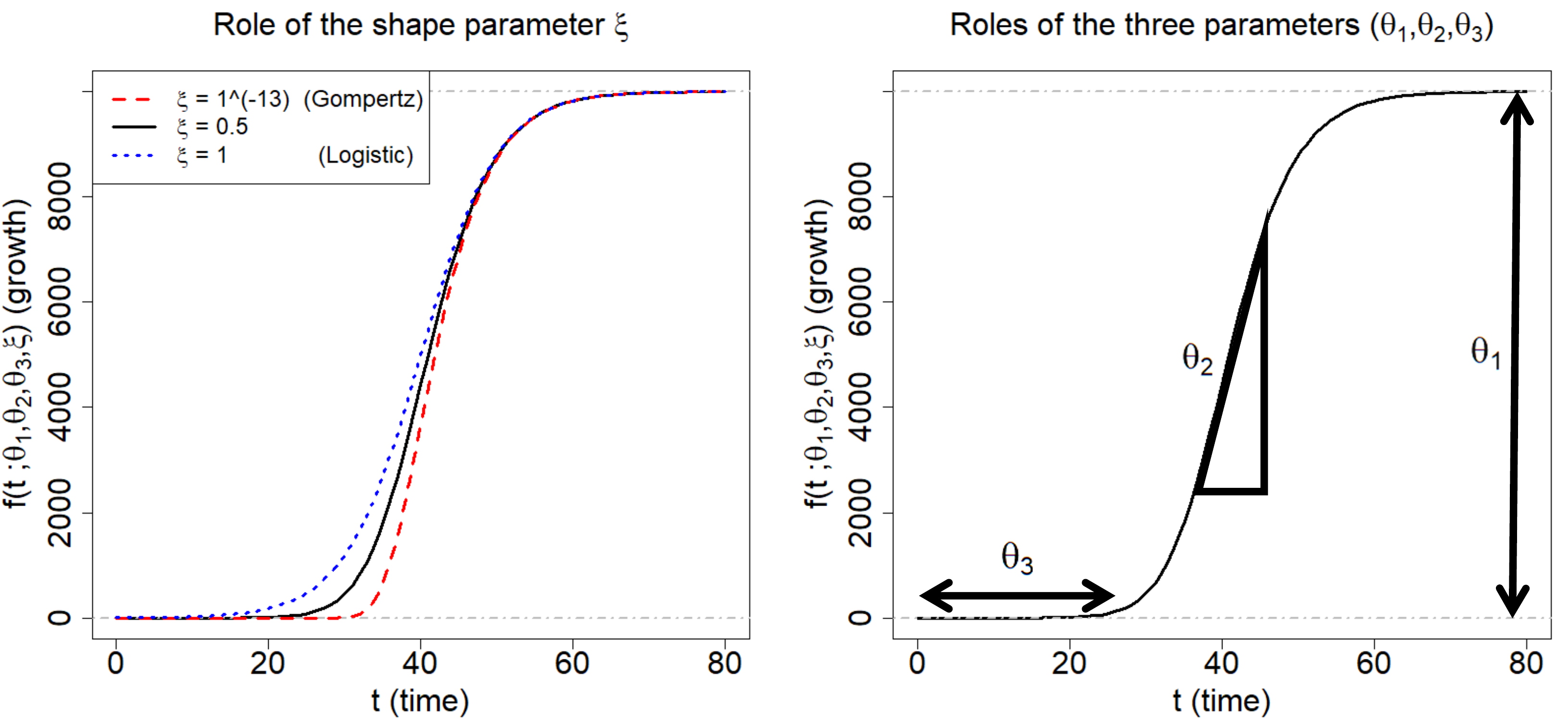}
    \caption{\baselineskip=10pt Description of the Richards growth curve model. The curve is obtained when $(\theta_1,\theta_2,\theta_3) = (10000,0.2,40)$. The left panel is obtained by changing the $\xi$ to be $1 \times 10^{-13}$, $0.5$, and $1$, respectively. The right panel describes the roles the three parameters in epidemiological modeling: $\theta_1$ represents final epidemic size; $\theta_2$ is an infection rate; and $\theta_3$ is a lag phase.}
    \label{fig:Richards curve}
\end{figure}

Figure \ref{fig:Richards curve} illustrates roles of the four parameters of the Richards curve (\ref{eq:Richards growth curve}). The curves on left panel is obtained when $(\theta_1,\theta_2,\theta_3) = (10000,0.2,40)$, while varying the $\xi$ to be $1 \times 10^{-13} (\approx 0)$, $0.5$, and $1$, respectively. The right panel pictorially describes the roles of $(\theta_1,\theta_2,\theta_3)$: $\theta_1$ represents the asymptote of the curve; $\theta_2$ is related to a growth rate (analytically, the derivative of logarithm of the curve (\ref{eq:Richards growth curve}) at $t=\theta_3$ is $\theta_2/2$.); and $\theta_3$ sets the displacement along the x-axis. (For more technical detail for the parameters, refer to \citep{birch1999new}.)

In epidemiological modeling, the Richards curve (\ref{eq:Richards growth curve}) can be used as a parametric curve describing infection trajectories shown in the Figure \ref{fig:6_countries}. In this context, each of the parameters can be interpreted as follows: $\theta_1$ represents the final epidemic size (that is, the maximum cumulative number of infected cases across the times);  $\theta_2$ represents infection rate; and  $\theta_3$ represents a lag phase of the trajectory. (The shape parameter $\xi$ seems to have no clear epidemiological meaning \citep{wang2012richards}.)

%%%%%%%%%%%%%%%%%%%%%%%%%%%%%%%%%%%
\subsection{Bayesian hierarchical Richards model}\label{subsec:Bayesian hierarchical model}
We propose a Bayesian hierarchical model based on the Richards curve (\ref{eq:Richards growth curve}), which is referred to as Bayesian hierarchical Richards model (BHRM), to accommodate the COVID-19 data $\{\textbf{y}_i,\textbf{x}_i \}_{i=1}^{N}$. (Although the model is based on the Richards curve, the idea can be generalized to any choice for growth curves.) Ultimately, a principal goal of the BHRM is to establish two functionalities:
\begin{itemize}
\baselineskip=15pt
\item[(a)] [Extrapolation] uncover a hidden pattern from the infection trajectory for each country $i$, that is, $\textbf{y}_{i} = (y_{i1},\cdots,y_{iT})^{\top}$, through the Richards growth curve $f(t ; \theta_1, \theta_2, \theta_3, \xi)$ (\ref{eq:Richards growth curve}), and then extrapolate the curve.
\item[(b)] [Covariates analysis] identify important predictors among the $p$ predictors $\textbf{x} = (x_1, \cdots, x_p )^{\top}$ that largely affect on the shape the curve $f(t ; \theta_1, \theta_2, \theta_3, \xi)$ in terms of the three curve parameters.
\end{itemize}
A hierarchical formulation of the BHRM is given as follows.
First, we introduce an additive independently identical Gaussian error to each observation $\{y_{it}\}_{i=1,t=1}^{N,T}$, leading to a likelihood part:
\begin{align}
\label{eq:likelihood_part}
y_{it}&=f(t ; \theta_{1i}, \theta_{2i}, \theta_{3i}, \xi_i) + \epsilon_{it}, \quad \epsilon_{it}\sim \mathcal{N}(0,\sigma^2),\,\,\,\,\,\,\,\,\,\,\,\,\,\,\,\,\, (i=1, \cdots, N, \, t = 1, \cdots, T),
\end{align}
where $f(t ; \theta_{1i}, \theta_{2i}, \theta_{3i}, \xi_i)$ is the Richards growth curve (\ref{eq:Richards growth curve}) which describes a growth pattern of infection trajectory for the $i$-th country. Because each of the curve parameters $(\theta_{1},\theta_{2},\theta_{3})$ has its own epidemiological interpretations, we construct three separate linear regressions:
\begin{align}
\label{eq:linear_regression}
\theta_{li} &= \alpha_l + \textbf{x}_i^{\top} \bm{\beta}_l + \varepsilon_{li}, \quad \varepsilon_{li}\sim \mathcal{N}(0,\sigma_l^2),\,\,\,\,\,\,\,\,\,\,\,\,\,\,\,\,\,\,\,\,\,\,\,\,\,\,\,\,\,\, (i=1, \cdots, N, \, l = 1, 2,3),
\end{align}
where $\bm{\beta}_l=(\beta_{l1},\cdots,\beta_{lj},\cdots, \beta_{lp})^{\top}$ is a $p$-dimensional coefficient vector corresponding to the $l$-th linear regression.

For the shape parameter $\xi$, we assume the standard log-normal prior:
\begin{align}
\label{eq:xi_log_normal}
    \xi_i &\sim \log\ \mathcal{N}(0,1),\,\,\,\,\,\,\,\,\,\,\,\,\,\,\, (i=1, \cdots, N).
\end{align}
The motivation of choosing the log-normal prior (\ref{eq:xi_log_normal}) for the $\xi_i$ is that the prior puts effectively enough mass on the region $(0,3)$ where most of the estimates for the $\xi_i$ ($i=1,\cdots,N$) concentrated on. Additionally, Gaussianity prior assumption makes it possible to employ the elliptical slice sampler \citep{murray2010elliptical} in sampling from the full conditional posterior distribution of the $\xi_i$.

To impose a continuous shrinkage effect \citep{bhadra2019lasso} on each of the coefficient vectors, we adopt to use the horseshoe prior \citep{carvalho2009handling,carvalho2010}:
\begin{align}
\label{eq:Horseshoe}
\beta_{lj}&|\lambda_{lj},\tau_{lj}, \sigma_l^2  \sim \mathcal{N}(0, \sigma_l^2 \tau_l^2 \lambda_{lj}^2)
,\quad
\lambda_{lj},\tau_{lj} \sim \mathcal{C}^{+}(0,1),\,\,\,\,\,\,\,
(l = 1, 2,3,\, j = 1,\cdots, p).
\end{align}
Finally, improper priors \citep{gelman2004bayesian} are used for the intercept terms and error variances terms in the model:
\begin{align}
\label{eq:improper_priors}
\alpha_l \sim \pi(\alpha) \propto 1,\,\quad
\sigma^2 &,\sigma_l^2 \sim \pi(\sigma^2) \propto 1/\sigma^2,
\quad\quad\quad\quad (l = 1, 2,3).
\end{align}
Generally speaking, modeling framework of the BHRM (\ref{eq:likelihood_part}) -- (\ref{eq:improper_priors}) is widely called the \emph{nonlinear mixed effects model} or \emph{hierarchical nonlinear model}, a standard framework for analysis of data in the form of continuous repeated measurements over time on each individual from a sample of individuals drawn from a population of interest \citep{davidian1995nonlinear}. See SI Appendix for a posterior computation for the BHRM (\ref{eq:likelihood_part}) -- (\ref{eq:improper_priors}).
\section{Results}\label{sec:Results}
\subsection{Benefits from the information borrowing}\label{subsec:Information borrowing in forecasting}
We investigate the predictive performance of three Bayesian models based on the Richards growth curve. We start with the individual country-based model ($\mathcal{M}_1$) which has been widely used in the literature, reflecting the belief that individual countries are “unrelated.” Next, we extend the previous model to a hierarchical model by utilizing the infection trajectories of all the 40 countries ($\mathcal{M}_2$).  A limitation of $\mathcal{M}_2$ is that it lacks certain countrywide adjustments in estimating the trajectories. Next, we further upgrade this model by adding country-specific covariates in a hierarchical fashion  ($\mathcal{M}_3$). (For technical description for the three models, see S2 Appendix in the Supporting information .) Eventually, borrowing information across the 40 countries takes place in these two hierarchical models, $\mathcal{M}_2$ and $\mathcal{M}_3$, but not in the individual country-based model $\mathcal{M}_1$.

For evaluation criteria, we calculate the mean squared error (MSE) \citep{wasserman2013all} associated with the extrapolated infection trajectory for each of the 40 countries. Training and test data are designated as follows: given that $\textbf{y}_k = (y_{k,1},\cdots,y_{k,T})^{\top}$ is an infection trajectory of the $k$-th country spanning for $T$ days since January 22nd, and $d$ is the chosen test-day, then (i) the training data is set by the trajectory spanning for $T-d$ days since January 22nd (that is, $(y_{k,1}, \cdots, y_{k,T-d})$), and (ii) the test data is set by the $d$ recent observations (that is, $(y_{k,T-d+1}, \cdots, y_{k,T})$). 

For the two hierarchical models $\mathcal{M}_2$ and $\mathcal{M}_3$, the MSE is averaged over the 40 countries:
$$\text{MSE}_{d} 
= 
\frac{1}{40 \cdot d}
\sum_{k=1}^{40}
\sum_{r=T-d+1}^{T}
(
y_{k,r} - y_{k,r}^{*}
)^2,$$ where $y_{k,r}$ is the actual value for the cumulative confirmed cases of the $k$-th country at the $r$-th time point, and $y_{k,r}^{*}$ is the forecast value: more concretely, $y_{k,r}^{*}$ is the posterior predictive mean given the information from 40 countries. For the non-hierarchical model $\mathcal{M}_1$, the $y_{k,r}^{*}$ in the $\text{MSE}_{d}$ is acquired by using the data from the $k$-th country .

\begin{figure}[h]
    \centering
    \includegraphics[scale = 0.32]{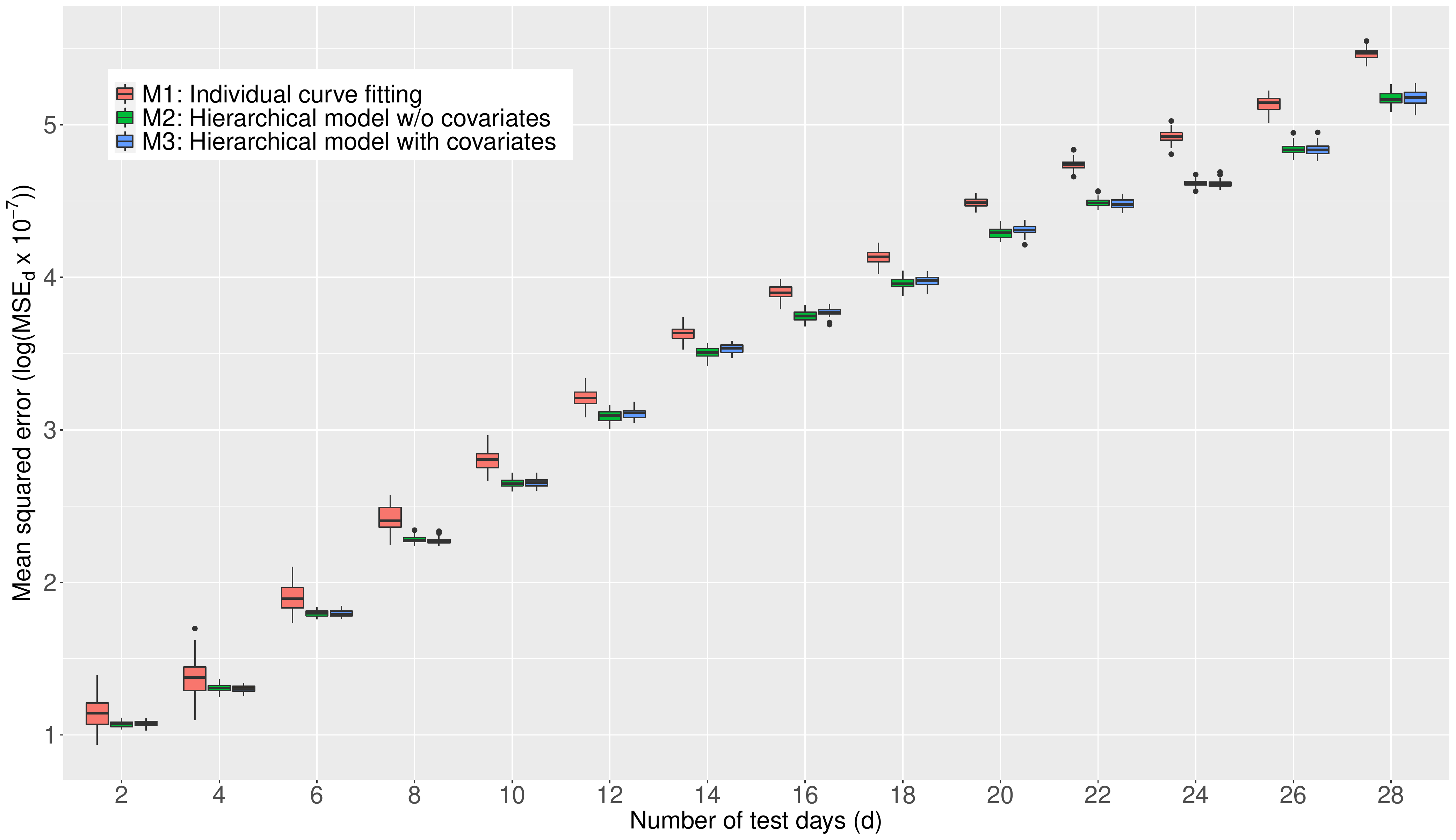}
    \caption{\baselineskip=10pt  Comparison of the MSE obtained by the three models, $\mathcal{M}_1$, $\mathcal{M}_2$, and $\mathcal{M}_3$, averaged over the $40$ countries. A smaller value for the MSE indicates a better predictive performance.}
    \label{fig:model_compare}
\end{figure}

For each of the test-days ($d = 2, 4, \cdots, 28$), we report the $\text{MSE}_d$'s from 50 replicates by showing the box plot. (The box plot \citep{boxplot} displays the distribution of $\text{MSE}_d$ fore each $\text{d}$ and model. The interquartile range (IQR) is represented as a box, which is from 25th percentile (Q1) to 75th percentile (Q3). A horizontal line in the box corresponds to the median. The maximun and minimun are set as $Q3 + 1.5*IQR$ and $Q1 - 1.5*IQR$, respectively. Any point larger than the maximun or smaller than the minumun is regarded as an outlier and drew as a black point in the plot.) The results are shown in Figure \ref{fig:model_compare}. From the panel, we see that (1) the predictive performances of two hierarchical models, $\mathcal{M}_2$ and $\mathcal{M}_3$, are better than that of $\mathcal{M}_1$ across the test-days; (2) the differences in the predictive performance between the non-hierarchical ($\mathcal{M}_1$) and the hierarchical models ($\mathcal{M}_2$ and $\mathcal{M}_3$) tend to get larger as the test-days increase; and (3) the predictive performances of two hierarchical models ($\mathcal{M}_2$ and $\mathcal{M}_3$) are similar across the test-days. Based on the outcomes, we shall conclude that information borrowing has improved the predictive accuracy in terms of MSE. A similar result where information borrowing is a benefit in improving predictive accuracy is found in the \emph{Clemente problem} from \citep{efron2010future} where the James-Stein estimator \citep{james1992estimation} better predicts then an individual hitter-based estimator in terms of the total squared prediction error. In what follows, we present all the results in the consequent subsections based on the model $\mathcal{M}_3$. 
\subsection{COVID-19 travel recommendations by country}\label{subsec:Maximum cumulative number of infected cases}
Centers for Disease Control and Prevention (CDC) categorizes countries into three levels by assessing the risk of COVID-19 transmission, used in travel recommendations by country (Visit \href{https://www.cdc.gov/coronavirus/2019-ncov/travelers/map-and-travel-notices.html}{www.cdc.gov/-}): Level 1, Level 2, and Level 3 indicate the Watch Level (Practice Usual Precautions), Alert Level (Practice Enhanced Precautions), and Warning Level (Avoid Nonessential Travel), respectively. 

We categorize the 40 countries into the three levels according to their posterior means for the final epidemic size (that is, $\theta_1$ of the Richards curve (\ref{eq:Richards growth curve})). Grouping criteria are as follows: (1) Level 1 (estimated total number is no more than 10,000 cases); (2) Level 2 (estimated total number is between 10,000 and 100,000 cases); and (3) Level 3 (estimated total number is more than 100,000 cases).

\begin{figure}[h]
    \centering
    \includegraphics[scale=0.45]{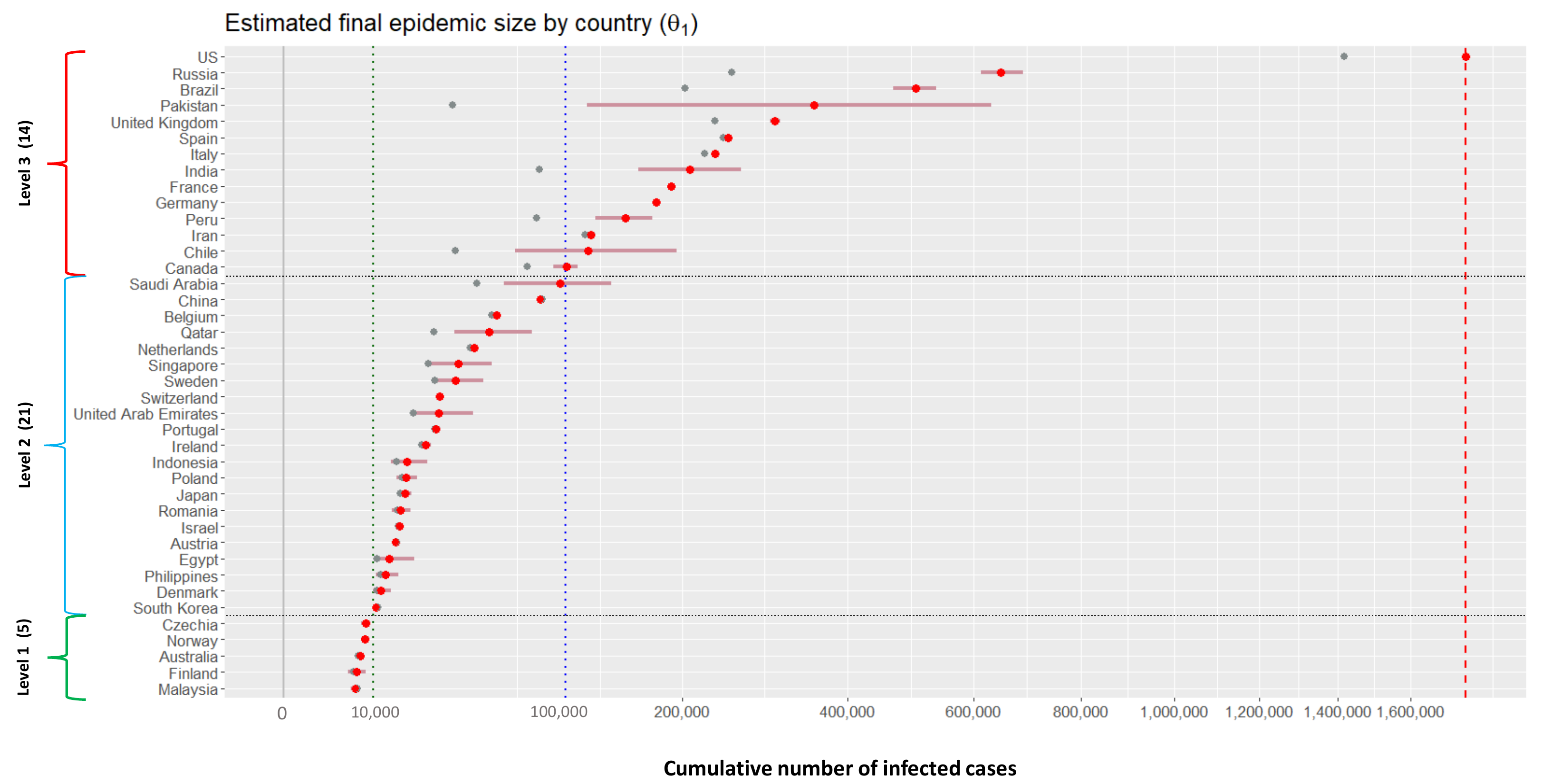}
    \caption{\baselineskip=10pt  Estimation results for the final epidemic size for 40 countries. Grey dots (\textcolor{gray}{$\bullet$}) represent the cumulative numbers of infected cases for 40 countries on May 14th; red dots (\textcolor{red}{$\bullet$}) and horizontal bars (\textcolor{red}{$-$}) represent the posterior means and $95\%$ credible intervals for the $\theta_1$ of the 40 countries. Vertical red dotted line indicates the $1,760,569$ cases, the posterior mean for the US.}
    \label{fig:maximums}
\end{figure}

Figure \ref{fig:maximums} displays results of posterior inference for the $\theta_1$ by country, based on the model $\mathcal{M}_3$. Countries on the $y$-axis are ordered from the severest country (US) to the least severe country (Malaysia) in the magnitude of the posterior means. The red horizontal bars on the panel represent the $95\%$ credible intervals describing the uncertainty regarding the estimated final epidemic size $\theta_1$. (Technically, lower and upper bounds of each of the intervals are obtained by taking the $2.5$-th and $97.5$-th percentiles from posterior samples of $\theta_1$ for the corresponding country, respectively.) Based on the results, there are 14 countries categorized as Level 3 (US, Russia, Brazil, Pakistan, UK, Spain, Italy, India, France, Germany, Peru, Iran, Chile, and Canada). There are 21 countries categorized as Level 2 (from Saudi Arabia to South Korea), and 5 countries categorized as Level 1 (from Czechia to Malaysia).

\subsection{Extrapolated infection trajectories and flat time points}\label{subsec:Extrapolation}
\begin{figure}[h]
    \centering
    \includegraphics[scale = 0.42]{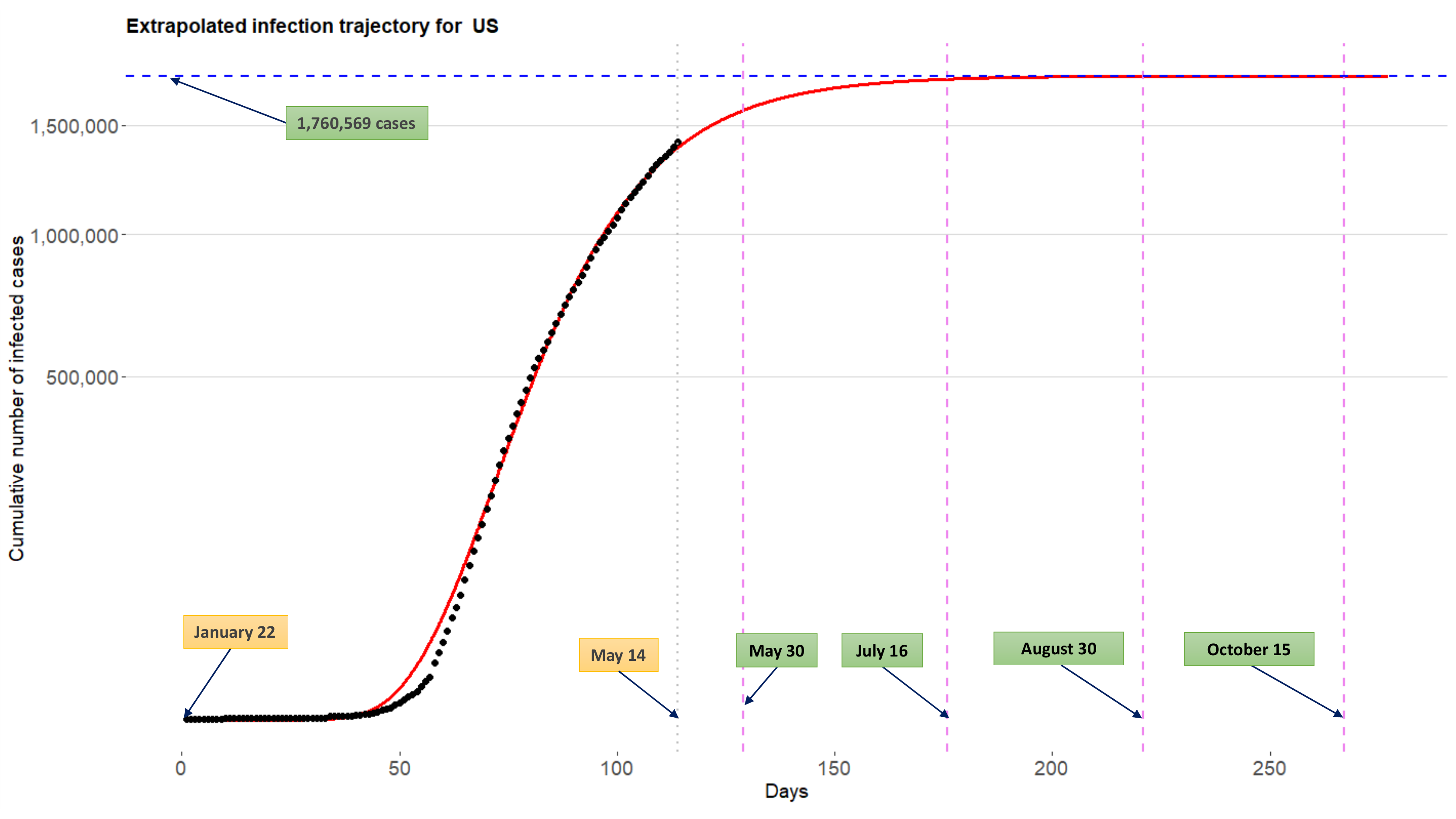}
    \caption{\baselineskip=10pt   Extrapolated infection trajectory for the US based on the model $\mathcal{M}_3$. Posterior mean of the maximum number of cumulative infected cases is is 1,760,569 cases. Posterior means for the flat time points are $t_{\text{flat},\gamma=0.9}$=May 30th, $t_{\text{flat},\gamma=0.99}$=July 16th, $t_{\text{flat},\gamma=0.999}$=August 30th, and $t_{\text{flat},\gamma=0.9999}$=October 15th.}
    \label{fig:US_it}
\end{figure}

Figure \ref{fig:US_it} displays the extrapolated infection trajectory (posterior mean for the Richards curve (\ref{eq:Richards growth curve})) for the US. The posterior mean of the final epidemic size is 1,760,569 cases. The scenario that `millions' of Americans could be infected was also warned by a leading expert in infectious diseases (Visit a related news article \href{https://www.bbc.com/news/world-us-canada-52086378}{www.bbc.com/-}). It is known that prediction of an epidemic trend from limited data during early stages of the epidemic is often futile and misleading \citep{hsieh2004sars}. Nevertheless, estimation of a possible severity havocked by the COVID-19 outbreak is an important task when considering the seriousness of the current pandemic situation.

A crucial question is when this trajectory gets flattened. To that end, we approximate a time point where at an infection trajectory levels off its value, showing a flattening pattern after the time point. The following is the definition of the \emph{flat time point} which we use in this paper: 

\begin{definition} Consider the Richards curve $f(t ; \theta_1, \theta_2, \theta_3, \xi)$ (\ref{eq:Richards growth curve}). Given a progression constant $\gamma\, (\%)$ with $0<\gamma<1$, the \emph{flat time point} $t_{\text{flat},\gamma}$ is defined as the solution of the equation
\begin{align}
\label{eq:flattening_progress}
    \gamma &= \frac{f(t ; \theta_1, \theta_2, \theta_3, \xi)}{\theta_1}\,\, (\%).
\end{align}
\end{definition}
By using elementary calculus, we can obtain the solution of the equation (\ref{eq:flattening_progress}):
\begin{align}
\label{eq:falt_time_point}
    t_{\text{flat},\gamma} &= \theta_3 - \frac{1}{\theta_2}\cdot
    \log \bigg[ \frac{1}{\xi} \cdot \bigg\{ \bigg(\frac{1}{\gamma}\bigg)^{\xi} -1 \bigg\}   \bigg]
    , \quad 0<\gamma<1.
\end{align}

\begin{figure}[h]
    \centering
    \includegraphics[scale=0.42]{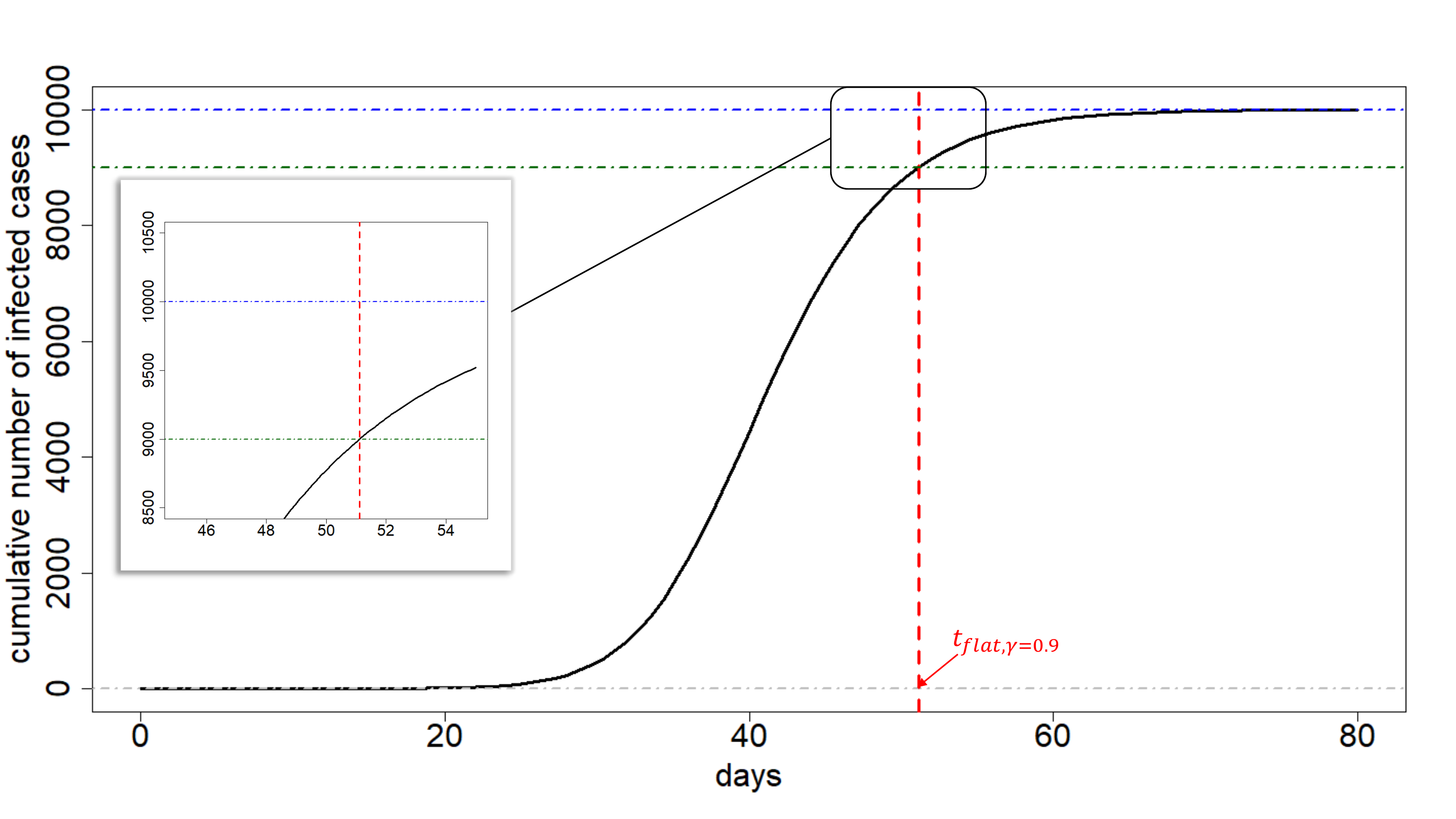}
    \caption{\baselineskip=10pt Illustration of flat time point. The exemplary infection trajectory is obtained by the Richards curve when $(\theta_1,\theta_2,\theta_3,\xi)=(10000,0.2,40,0.5)$. A flat time point $t_{\text{flat},\gamma}$ is approximately $51$ (vertical red dashed line). The vertical difference between the $\theta_1$ and the function value of Richards curve evaluated at $t_{\text{flat},\gamma}$ is $\gamma=0.9$.}
    \label{fig:flat_point}
\end{figure}

Technically, the flat time point $t_{\text{flat},\gamma}$ (\ref{eq:flattening_progress}) is interpreted as follow. Given a progression constant $\gamma\,\%$ (set by epidemiologist), the flat time point $t_{\text{flat},\gamma}$ is the time point whereat only $(1-\gamma) \theta_1$ cases can maximally take place to reach the final epidemic size $\theta_1$ following the time point $t_{\text{flat},\gamma}$. Here, the progression constant $\gamma \, (\%)$ is a value indicating a development of the pandemic: a higher value of $\gamma$ implies a later stage of the pandemic where at infection trajectories begin or tend to reach plateau. Figure \ref{fig:flat_point} depicts an exemplary infection trajectory based on the Richards curve (\ref{eq:Richards growth curve}) where the parameters were chosen by  $(\theta_1,\theta_2,\theta_3,\xi)=(10000,0.2,40,0.5)$, while the progression constant is set by $\gamma =0.9$, leading to flat time point $t_{\text{flat},\gamma}$ to be approximately $51$.

Standard choices for $\gamma$ shall be $0.9$, $0.99$, $0.999$, etc, because infection trajectories typically begins to display its flattening phase when $\gamma$ is equal to or greater than $0.9$. Choice of value for $\gamma$ depends on the particular situation of a country considered: for example, for China which already shows flattened trajectory (refer to Figure \ref{fig:6_countries}), $\gamma=0.999$ can be safely used, but for US one may use each of the $\gamma=0.9, 0.99, 0.999$, and $0.9999$ to further investigate evolvement of flattening phase over time.

For the US, the posterior means of the flat time points $t_{\text{flat},\gamma}$ are May 30th, July 16th, August 30th, and October 15th when corresponding $\gamma$'s are chosen by 0.9, 0.99, 0.999, and 0.9999, respectively. It is important to emphasize that the extrapolated infection trajectory is \emph{real-time prediction} of COVID-19 outbreaks \citep{fineberg2009epidemic,wang2012richards} based on observations tracked until May 14th. Certainly, incorporation of new information such as compliance with social distancing or advances in medical and biological sciences for this disease will change the inference outcomes.

\begin{figure}[H]
    \centering
    \includegraphics[scale = 0.35]{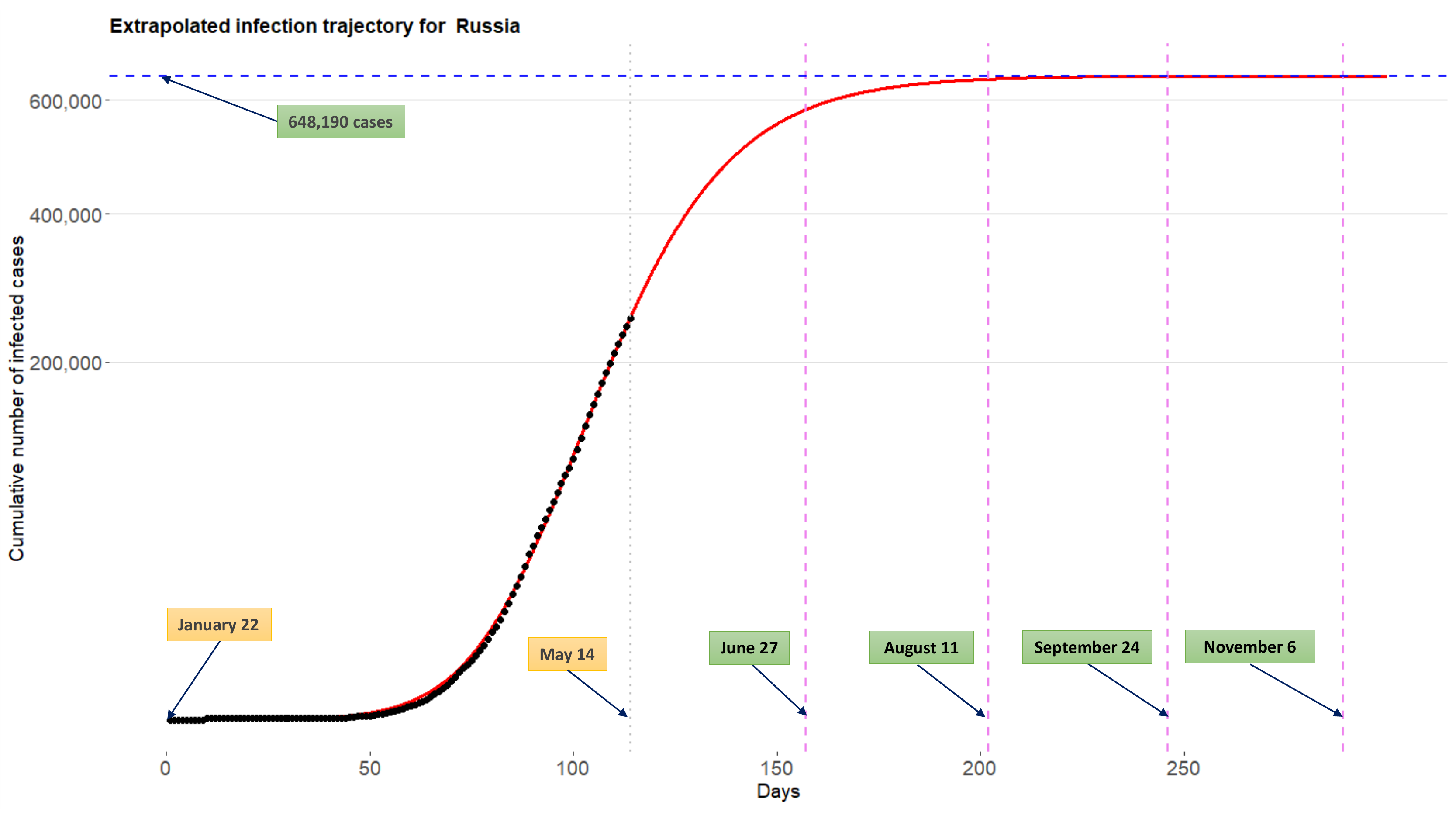}
    \includegraphics[scale = 0.35]{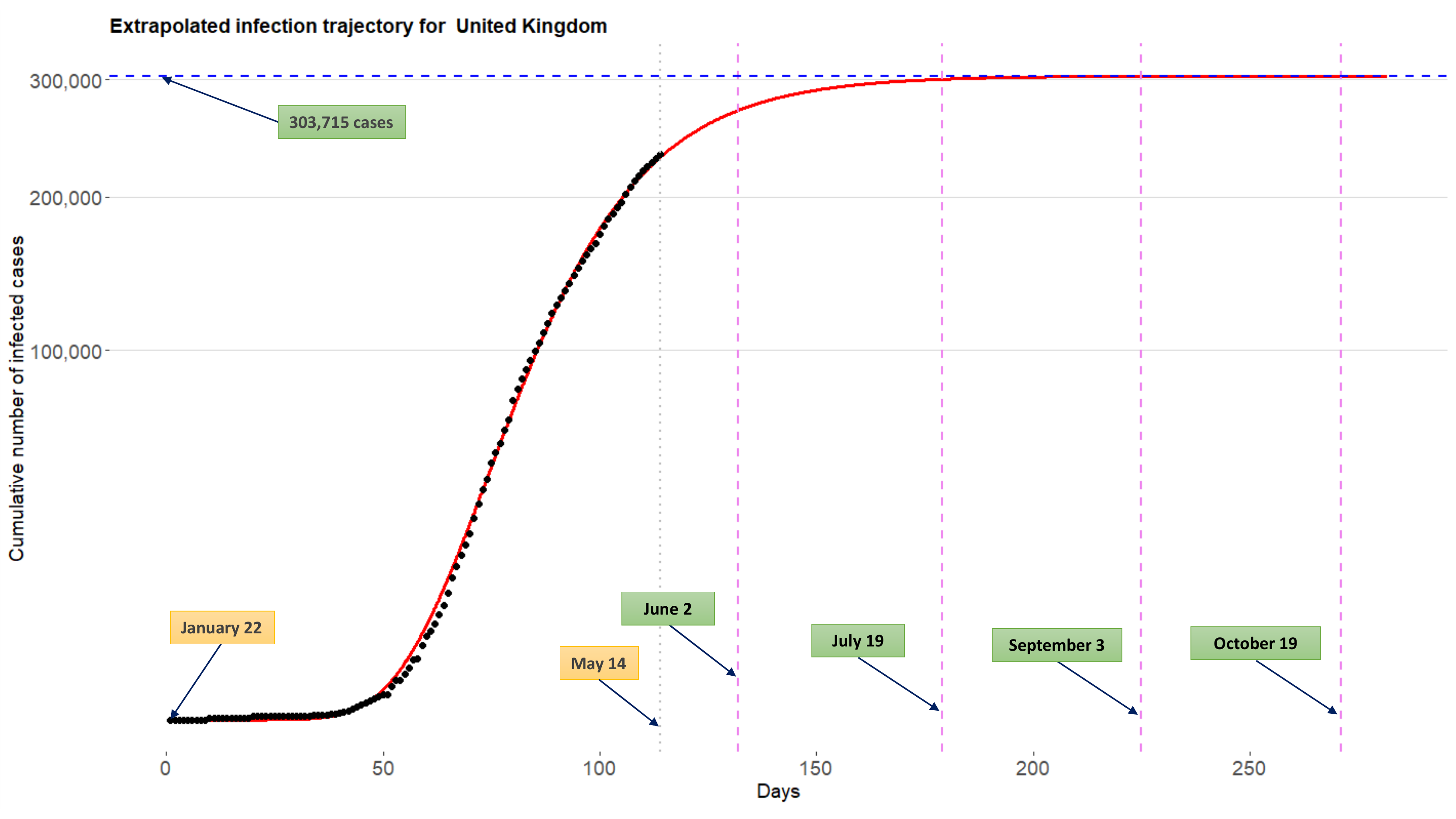}
    \includegraphics[scale = 0.35]{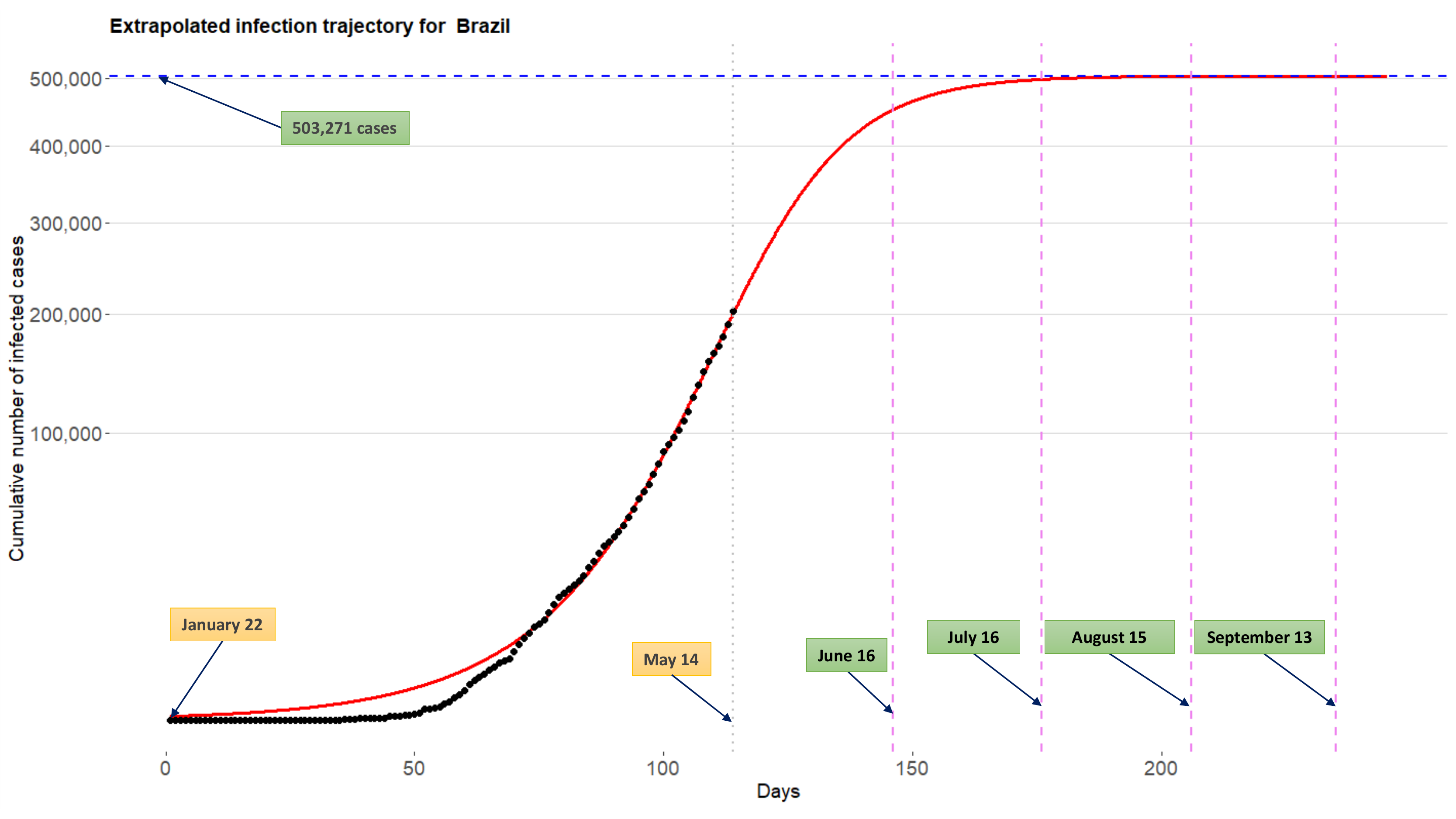}
    \caption{\baselineskip=10pt Extrapolated infection trajectory for the Russia (top), UK (middle), and Brazil (bottom). Flat time points are estimated by: (1) for the Russia, $t_{\text{flat},\gamma=0.9}$=June 27th, $t_{\text{flat},\gamma=0.99}$=August 11th, $t_{\text{flat},\gamma=0.999}$=September 24th, and $t_{\text{flat},\gamma=0.9999}$=November 6th; (2) for the UK, $t_{\text{flat},\gamma=0.9}$=June 2nd, $t_{\text{flat},\gamma=0.99}$=July 19th, $t_{\text{flat},\gamma=0.999}$=September 3rd, and $t_{\text{flat},\gamma=0.9999}$=October 19th; and (3) for the Brazil, $t_{\text{flat},\gamma=0.9}$=June 16th, $t_{\text{flat},\gamma=0.99}$=July 16th, $t_{\text{flat},\gamma=0.999}$=August 15th, and $t_{\text{flat},\gamma=0.9999}$=September 13th.}
    \label{fig:3_countries}
\end{figure}

 Figure \ref{fig:3_countries} shows the extrapolated infection trajectories for Russia, UK, and Brazil. Posterior means of the final epidemic size are as follows: (1) for the Russia, 648,190 cases; (2) for the UK, 303,715 cases; and (3) for the Brazil, 503,271 cases. Flat time points are estimated by: (1) for the Russia, $t_{\text{flat},\gamma=0.9}$=June 27th, $t_{\text{flat},\gamma=0.99}$=August 11th, $t_{\text{flat},\gamma=0.999}$=September 24th, and $t_{\text{flat},\gamma=0.9999}$=November 6th; (2) for the UK, $t_{\text{flat},\gamma=0.9}$=June 2nd, $t_{\text{flat},\gamma=0.99}$=July 19th, $t_{\text{flat},\gamma=0.999}$=September 3rd, and $t_{\text{flat},\gamma=0.9999}$=October 19th; and (3) for the Brazil, $t_{\text{flat},\gamma=0.9}$=June 16th, $t_{\text{flat},\gamma=0.99}$=July 16th, $t_{\text{flat},\gamma=0.999}$=August 15th, and $t_{\text{flat},\gamma=0.9999}$=September 13th. Results for other countries are included in the SI Appendix.

\subsection{Global trend for the COVID-19 outbreak}\label{subsec:Global trend for the COVID-19}
\begin{figure}[H]
    \centering
    \includegraphics[scale = 0.40]{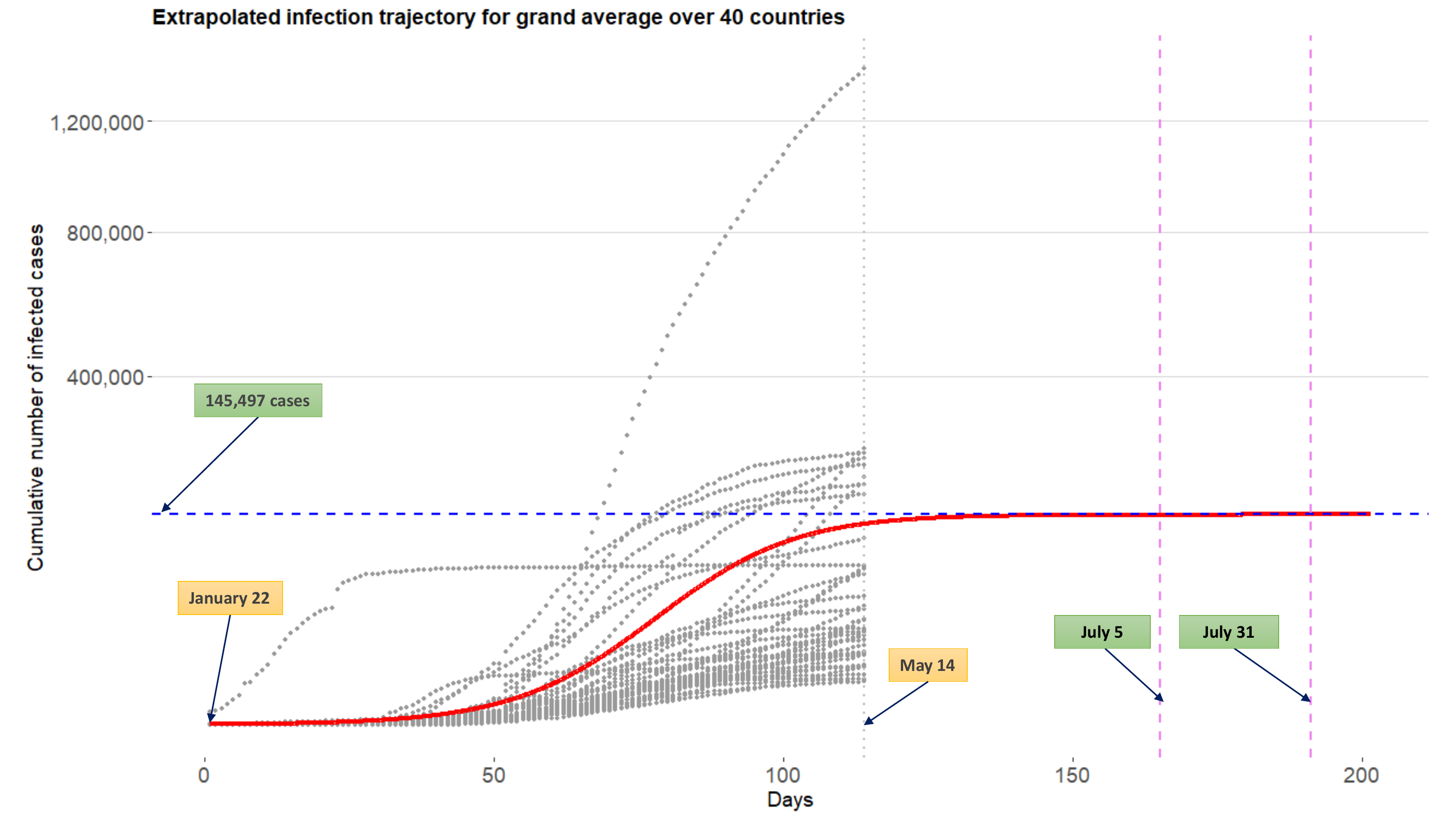}
    \caption{\baselineskip=10pt   Extrapolated infection trajectory for grand average over 40 countries obtained from the model $\mathcal{M}_3$. Grey dots are historical infection trajectories for 40 countries spanning from January 22nd to May 14th. Posterior means for the flat time points are $t_{\text{flat},\gamma=0.999}$=July 5th and $t_{\text{flat},\gamma=0.9999}$=July 31st.}
    \label{fig:grand_average}
\end{figure}

Figure \ref{fig:grand_average} displays the extrapolated infection trajectory for grand average over 40 countries obtained from the model $\mathcal{M}_3$. Technically, this curve is acquired by extrapolating the Richards curve by using the intercept terms in linear regressions (\ref{eq:linear_regression}). The grey dots on the panel are historical infection trajectories for 40 countries. Posterior means for the final epidemic size is 145,497 cases. Posterior means for the flat time points are $t_{\text{flat},\gamma=0.999}$=July 5th and $t_{\text{flat},\gamma=0.9999}$=July 31st.

\subsection{Identifying risk factors for severe disease due to COVID-19}\label{subsec:Regression analysis}
COVID-19 is a new disease and there is very limited information regarding risk factors for this severe disease. There is no vaccine aimed to prevent the transmission of the disease because there is no specific antiviral agent is available. It is very important to find risk factors relevant to the disease. Reliable and early risk assessment of a developing infectious disease outbreak allow policymakers to make swift and well-informed decisions that would be needed to ensure epidemic control.

CDC described High-Risk Conditions based on currently available information and clinical expertise (For more detail, visit  \href{https://www.cdc.gov/coronavirus/2019-ncov/hcp/underlying-conditions.html}{www.cdc.gov/-}): those at higher risk for infection, severe illness, and poorer outcomes from COVID-19 include
\begin{itemize}
\baselineskip=15pt
    \item People 65 years and older;
    \item People who live in a nursing home or long-term care facility;
    \item People with chronic lung disease or moderate to severe asthma;
    \item People who are immunocompromised, possibly caused by cancer treatment, smoking, bone marrow or organ transplantation, immune deficiencies, poorly controlled HIV or AIDS, and prolonged use of corticosteroids and other immune weakening medications;
    \item People with severe obesity (body mass index of 40 or higher);
    \item People with diabetes; 
    \item People with chronic kidney disease undergoing dialysis;
    \item People with liver disease.
\end{itemize}

The model $\mathcal{M}_3$ involves three separated linear regressions indexed by $l=1,2,$ and $3$, whose response and coefficient vector are denoted by $\theta_l$ and $\bm{\beta}_l$, respectively ($l=1,2,3$). (See the equation  (\ref{eq:linear_regression})) The sparse horseshoe prior \citep{carvalho2009handling,carvalho2010} is imposed for each of the coefficient vectors, which makes the model equipped with covariates analysis. That way, we can identify important predictors explaining the heterogeneity of shapes existing in infection trajectories across 40 countries. Because each of the responses has its own epidemiological interpretation (final epidemic size ($\theta_1$), infection rate ($\theta_2$), and lag phase ($\theta_3$)), the joint variable selection techniques employed by the sparse horseshoe prior can be further used in finding possible risk factors for severe disease due to COVID-19 among the 45 predictor considered in this research.

Table \ref{table:top_10_covariates} summarizes 10 significant predictors among the $45$ predictors, explaining each of the responses $\theta_l$ ($l=1,2,3$). Contents of the table are listed with the form ``covariate name (estimate of coefficient)". Technically, the estimate inside of the parenthesis is the posterior mean for the coefficient. By following convention in variable selection schemes as done by several authors \citep{tibshirani1996regression,armagan2013generalized}, we standardized the design matrix and then make a posterior inference: therefore, an estimated coefficient may not indicate a change of value for the response $\theta_l$ ($l=1,2,3$) with respect to a unit increment of value for covariate. Rather, the estimate can be thought of as a measure representing the sensitiveness of a change in response with respect to a change in the covariate.
\begin{table}[ht]
\begin{scriptsize}
\small
\centering
\caption{\baselineskip=10pt
Important predictors explaining $\theta_{l}$, $l=1,2,3$}
 \label{table:top_10_covariates}
\begin{tabular}{lll}
\hline
\hline
\multicolumn{1}{c}{Final epidemic size ($\theta_1$)}       & \multicolumn{1}{c}{Infection rate ($ \theta_2$)}       & \multicolumn{1}{c}{Lag phase ($\theta_3$)} \\ \hline
Insuf\_phy\_act($+$19895)              &     Points\_of\_Entry($-$0.0107)      & Dis\_to\_China($+$16.14)    \\
Testing\_num($+$19256)        &         Alcohol\_consumers\_total($+$0.0106)             &       Alcohol\_cons\_rec($-$5.24)          \\
Testing\_popu($-$16596)            &         Alcohol\_cons\_rec($+$0.0071)              &        Median\_age($-$4.70)         \\
Overweight($+$16173)          &         Air\_pollution($+$0.0055)             &     Alcohol\_consumers\_total($-$4.33)           \\
MCV1\_immun($-$15447)            &        Life\_expect\_total\_60($+$0.0054)             &      Testing\_num($-$4.16)          \\
Testing\_confirm($+$10689)             &     Popu\_density($+$0.0050)      & Life\_expect\_total\_60($-$1.93) \\
Pol3\_immun($-$8388)       &        Laboratory($-$0.0046)              &     Cigarette\_smoke($-$1.85)           \\
Hib3\_immun($-$7478)              &         Heavy\_drinking\_total($+$0.0044)              &      Tuberculosis\_case($-$1.80)          \\
Tempe\_avg($-$4573)       &        Cigarette\_smoke($+$, 0.0026)              &     Total\_over\_65($-$1.35)           \\
Alcohol\_cons\_rec($+$4205)              &         Tobacco\_smoke($+$0.0025)              &      Tobacco\_smoke($-$1.29)          \\
\hline
\end{tabular}
\begin{itemize}
    \item[] NOTE: the table shows 10 interesting covariates for each parameter. They are listed with ``covariate name (estimate of coefficient)" where the estimate is the posterior mean for the corresponding coefficient. See SI Appendix for a detailed explanation for the listed covariates.
\end{itemize}
\end{scriptsize}
\end{table}

The followings are general guidelines on how covariates on the Table \ref{table:top_10_covariates} can be interpreted in the current context of pandemic. 
\begin{itemize}
\baselineskip=15pt
    \item The parameter $\theta_{1}$ represents \emph{final epidemic size}. A larger number of $\theta_{1}$ indicates that a country has (can have) more COVID-19 infected patients in the country. A covariate with a positive estimate (or negative estimate) is a factor associated with an increase (or decrease) of the total infected cases. A covariate endowed with a larger magnitude (that is, absolute value) for the estimate makes a greater influence on $\theta_{1}$. 
    \item The parameter $\theta_{2}$ represents \emph{infection rate}. A larger number of $\theta_{2}$ implies a faster spread of the virus around the country. A covariate with a positive estimate (or negative estimate) is a factor associated with a rapid (or slow) spread of the virus. A covariate endowed with a larger magnitude for the estimate makes a greater influence on $\theta_{2}$.
    \item The parameter $\theta_{3}$ represents \emph{lag phase} of the infection trajectory. The larger the value of $\theta_3$ the later the trajectory begins to accumulate infected cases, leading to a later onset of the accumulation. A covariate with a positive estimate (or negative estimate) is a factor associated with delaying (or bring forward) the onset of the accumulation. A covariate endowed with a larger magnitude for the estimate makes a greater influence on $\theta_{3}$. 
\end{itemize}

%Its estimate is $+$19895; this implies that final epidemic size will increase by 19895 cases if the proportion of people without enough physical activity increases by 1\% on average (See SI Appendix for the unit for covariates.). 
%The average temperature is negatively related to the epidemic size. An increment of 1 degree Fahrenheit of temperature may reduce the final epidemic size by 4205 cases. 
%, where a delay of 16.14 days may happen if the distance increase by 1 unit

Now, based on the aforementioned guideline, we shall interpret the Table \ref{table:top_10_covariates} in detail. (The reasoning reflects our subjectivity, and disease expert should decipher precisely.)

As for the parameter $\theta_1$, insufficient physical activity has been selected as one of the important risk factors which may increase the final epidemic size of a country. Additionally, intense immunization coverage on measles, Polio, and Haemophilus Influenzae type B can reduce the final epidemic size. Poor general health status of a population \citep{Demographicscience} such as overweight and alcohol addiction can increase the epidemic size. (visit related news article \href{https://www.cidrap.umn.edu/news-perspective/2020/04/new-york-obesity-appears-raise-covid-19-risk}{www.cidrap.umn.edu/-}.) Certain testing information is also associated with the epidemic size, which can be further researched in retrospective studies in swift policymaking for a future pandemic. (See a WHO report for the relationship between climate change and infectious diseases \href{https://www.who.int/globalchange/climate/summary/en/index5.html}{www.who.int/-}.) 

Turning to the parameter $\theta_2$, a rigorous fulfillment of general obligations at point of entry is chosen as one of the significant predictors in reducing the infection rate. Additionally, poor smoking and alcoholic behaviors of a country population are risk factors that may increase the infection rate. Demographically, it has been found that densely populated countries or countries where life expectancy is relatively high are more venerable to the rapid disease transmission among people. Among national environmental status, poor air condition which may negatively influence people's respiratory system is found to be a risk factor increasing the infection rate. 

Finally, moving to the parameter $\theta_3$, a geological distance of a certain country from China is an important covariate delaying the onset of the infected cases. The lag of onset is also graphically observed from the Figure \ref{fig:6_countries}: time point whereat South Korea begins to accumulate the infected cases is relatively earlier than those of the US, UK, etc. Similar to $\theta_2$, heavier alcohol drinking and tobacco use may result in an earlier onset of the accumulation of the infected patients, thereby bringing forward the infection trajectory. Having larger numbers of median age and elderly people of a population can shorten the lag phase. Finally, conducting frequent testing for the COVID-19 helps detect infected patients, followed by the earlier accumulation for the confirmed cases.

\section{Discussions}\label{sec:Discussions}
In general, there are three major categories of infectious disease prediction models: (i) differential equation models, (ii) time series models, and (iii) the statistical models. The differential equation models describe the dynamic behavior of the disease through differential equations allowing the laws of transmission within the population. The popular models include the SI, SIS, SIR, and SEIR models \citep{2000SIAMR..42..599H, ExactanalSIR,korobeinikov2004lyapunov}. These models are based on assumptions related to S (susceptible), E (exposed), I (infected), and R (remove) categories of the population. Time series based prediction models such as ARIMA, Grey Model, Markov Chain models have been used to describe dependence structure over of the disease spread over time \citep{EpidemiologyARIMA,GMpredict, hu2006rainfall,rushton2006disease,GeneralizedMarkov}. On the other hand, statistical models, so-called phenomenological models, which follow certain laws of epidemiology \citep{clayton2013statistical,thompson2006epidemiology} are widely used in real-time forecasting for infection trajectory or size of epidemics in early stages of pandemic \citep{fineberg2009epidemic,hsieh2009richards,pell2018using}. Statistical models can be easily extended to the framework of hierarchical models (multilevel models \citep{snijders2011multilevel}) to analyze data within a nested hierarchy, eventually harnessing the data integration \citep{hill1965inference,tiao1965bayesian,stone1965paradox,browne2006comparison}. In this paper, we proposed a Bayesian hierarchical model, BHRM (\ref{eq:likelihood_part}) -- (\ref{eq:improper_priors}), so that data integration and uncertainty analysis \citep{malinverno2004expanded} are possible in a unified way. 

BHRM is a Bayesian version of a two-stage non-linear mixed effect model \citep{davidian1995nonlinear} where the first and second stages are related to the curve-fitting based on a certain parametric curve (we used Richards curve (\ref{eq:Richards growth curve})) and covariates analysis, respectively. In such a non-linear modeling framework, one of the challenges is an accommodation of individual-specific covariates which change over the course of observations. (See page 149 in \citep{davidian1995nonlinear} and \citep{fitzmaurice2008longitudinal} for more detailed discussion on this issue.) However, in infectious disease modeling for COVID-19 spread curves, there are important time-varying covariates that can be further considered in a possible model: examples include daily number of COVID-19 tests conducted, daily social distancing scores, etc. These time-varying covariates may be used in data level (likelihood level) endowed with center-adjusted inference for the curve-fitting \citep{li2010bayesian}.

It is important to point out that the real-time forecast during early stages of the pandemic may result in premature inference outcomes \citep{hsieh2004sars}, but it should not demoralize predictive analysis as the entire human race is currently threatened by unprecedented crisis due to COVID-19 pandemic. To improve the predictive accuracy, data integration from multiple countries was a key notion, which is closely related to borrowing information. The motivation of using the borrowing information is to make use of \emph{indirect evidence} \citep{efron2010future} to enhance the predictive performance: for example, to extrapolate the infection trajectory for the US, the information not only from the US (\emph{direct evidence}) but also from other countries (\emph{indirect evidence}) are utilized to better predict the trajectory for the US. Further, to render the information borrowing endowed with uncertainty quantification, Bayesian argument is inevitable, inducing sensible inferences and decisions for users \citep{lindley1972bayesian}.

\section{Conclusion}
It is important to emphasize that, while medical and biological sciences are on the front lines of beating back COVID-19, the true victory relies on advance and coalition of almost every academic field. However, information about COVID-19 is limited: there are currently no vaccines or other therapeutics approved by the US Food and Drug Administration to prevent or treat COVID-19 (on April 13, 2020). Although numerous research works are progressed by different academic fields, the information about COVID-19 is scattered around different disciplines, which truly requires interdisciplinary research to hold off the spread of the disease. 

In this paper, we proposed the BHRM (\ref{eq:likelihood_part}) -- (\ref{eq:improper_priors}) based on the Richards growth curve (\ref{eq:Richards growth curve}) \citep{richards1959flexible}. In summary, the novelties of our method are as follows: we (i) used a flexible hierarchical growth curve model to global COVID-19 data, (ii) integrated information from 40 countries for estimation and prediction purposes, and (iii) performed covariate analysis to find important reasons to explain the heterogeneity in the country-wise infection trajectories across 40 countries.

The results demonstrated the superiority of our approach compared to an existing individual country-based model. Our research outcomes can be thought even more insightful given that we have not employed information about disease-specific covariates. That being said, using more detailed information such as social mixing data, precise hospital records, or patient-specific information will further improve the performance of our model. Moreover, integration of epidemiological models with these statistical models will be our future topic of research. 

\clearpage
\baselineskip=14pt
\baselineskip=14pt
\bibliographystyle{chicago}
\bibliography{ref}

\setcounter{equation}{0}
\setcounter{table}{0}
\setcounter{figure}{0}
\renewcommand{\theequation}{S.\arabic{equation}}
\renewcommand{\thetable}{S.\arabic{table}}
\renewcommand{\thefigure}{S.\arabic{figure}}
\baselineskip=17pt

\section*{Supporting Information Appendix}

\section*{S1 Appendix: Research data}\label{S1_Appendix}
In this research, we analyze global COVID-19 data $\{\textbf{y}_i,\textbf{x}_i \}_{i=1}^{N}$, obtained from $N = 40$ countries. These 40 countries were selected for our analysis targets because cumulative numbers for the infected cases for the 40 countries were top 40 largest (hence, top 40 severest) in the world on the date May 14th, 2020. (Meanings for the vector notations, $\textbf{y}_i$ and $\textbf{x}_i$, will be explained shortly later.) These countries are most severely affected by the COVID-19 in terms of the confirmed cases on May 14th, and listed on Table \ref{table: 40 countries}: each country is contained in the table with format ``country name (identifier)", and this identifier also indicates a severity rank, where a lower value indicates a severer status. The order of the ranks thus coincides with the order of the countries named on the $y$-axis of the Figure \ref{fig:Confirmed cases on May 14th}.
\begin{table}[h]
\caption{\baselineskip=10pt  \ 40 countries on the research}
\centering
\begin{tabular}{l}
\hline
\hline
\multicolumn{1}{c}{Country (index $i$)} \\ \hline
US (1), Russia (2), Spain (3), United Kingdom (4),  Italy (5),\\
Brazil (6), France (7), Germany (8),  Iran (9),  China (10),\\
India (11), Peru (12), Canada (13), Belgium (14), Saudi Arabia (15),\\
Netherlands (16), Chile (17), Pakistan (18), Switzerland (19), Portugal (20),\\
Sweden (21), Qatar (22), Singapore (23), Ireland (24), United Arab Emirates (25), \\
Poland (26), Japan (27), Israel (28), Romania (29), Austria (30),\\
Indonesia (31), Philippines (32), South Korea (33), Denmark (34), Egypt (35), \\
Czechia (36), Norway (37), Australia (38), Malaysia (39), Finland (40)\\
\hline
\end{tabular}
\begin{itemize}
    \item[] NOTE: Countries are listed with the form ``country name (identifier)". This identifier also represents a severity rank. The rank is measured based on the accumulated number of the confirmed cases on May 14th.
\end{itemize}
\label{table: 40 countries}
\end{table}

For each country $i$ ($i=1,\cdots, N$), let $y_{it}$ denotes the number of accumulated confirmed cases for COVID-19 at the $t$-th time point ($t=1,\cdots,T$). Here, the time indices $t=1$ and $t=T$ correspond to the initial and end time points, January 22nd and May 14th, respectively, spanning for $T=114$ (days). The time series data $\textbf{y}_i = (y_{i1},\cdots,y_{it}, \cdots, y_{iT})^{\top}$ is referred to as an \emph{infection trajectory} for the country $i$. Infection trajectories for eight countries (US, Russia, UK, Brazil, Germany, China, India, and South Korea) indexed by $i=1$, $2$, $4$, $6$, $8$, $10$, $11$, and $33$, respectively, are displayed in the Figure \ref{fig:6_countries}. We collected the data from the Center for Systems Science and Engineering at the Johns Hopkins University.

For each country $i$, we collected 45 covariates, denoted by $\textbf{x}_i = (x_{i1},\cdots,x_{ij},  \cdots, x_{ip})^{\top}$ ($p=45$). The $45$ predictors can be further grouped by 6 categories: \emph{the 1st category}: general country and population distribution and statistics; \emph{the 2nd category}: general health care resources; \emph{the 3rd category}: tobacco and alcohol use; \emph{the 4th category}: disease and unhealthy prevalence; \emph{the 5th category}: testing and immunization statistics; and \emph{the 6th category}: international health regulations monitoring. The data sources are the World Bank Data (\href{https://data.worldbank.org/}{https://data.worldbank.org/-}), World Health Organization Data (\href{https://apps.who.int/gho/data/node.main}{https://apps.who.int/-}), and National Oceanic and Atmospheric Administration (\href{https://www.noaa.gov/}{https://www.noaa.gov/-}). Detailed explanations for the covariates are described in \nameref{Tables_for_covariates}.

\subsection*{Tables for covariates}\label{Tables_for_covariates}

\begin{table}[H]
\caption{\baselineskip=10pt  \ Category of covariates.}
\centering
\begin{tabular}{ll}
\hline
\hline
\multicolumn{1}{c}{Category} & \multicolumn{1}{c}{Covariates (index)} \\ \hline
General country and  & Total\_over\_65 (1), Female\_per (2), Median\_age (5), \\
population distribution & Birth\_rate (6), Life\_expect\_total\_60 (14), \\
and statistics & Dis\_to\_China (40), Popu\_density (44), Tempe\_avg (45)\\
Health care resources & Physician (3), Doc\_num\_per (12), Hosp\_bed (13)\\
Tobacco and alcohol use & Alcohol\_cons\_rec (7), Alcohol\_cons\_unrec (8), \\
& Alcohol\_consumers\_total (9), Heavy\_drinking\_total (10), \\
& Alcohol\_death\_total (11), Tobacco\_smoke (34), \\
& Cigarette\_smoke (35)\\
Disease and unhealth & Underweight\_total (4), Blood\_glucose (30), \\
prevalence & Cholesterol (31), Insuf\_phy\_act (32), Overweight (33), \\
& Air\_pollution (36), Air\_pollution\_death (37), \\
& Air\_pollution\_DALYs (38), Tuberculosis\_case (39), \\
Testing and immunization  & Dtt\_dtp\_immun (15), HepB3\_immun (16), Hib3\_immun (17),\\
statistics & MCV1\_immun (18), MCV2\_immun (19), PCV3\_immun (20),\\
& Pol3\_immun (21), Testing\_num (41), Testing\_confirm (42),\\
&  Testing\_popu (43)\\
International Health  & Zoonotic\_Events (22), Food\_Safety (23), Laboratory (24), \\
Regulations monitoring & Human\_Resources (25), Health\_Service\_Provision (26), \\
& Risk\_Communication (27), Points\_of\_Entry (28), \\
& Radiation\_Emergencies (29)\\ \hline
\end{tabular}
\begin{itemize}
    \item[] NOTE 1: Covariates are listed with the form ``predictor name (index)". Predictor names are abbreviated.
    \item[] NOTE 2: For part of the covariates, the corresponding data available for the public are not for the record of 2020, which may introduce some possible bias in the estimation.
\end{itemize}
\label{table: 45-predictors_main_body}
\end{table}

\begin{table}[h]
\caption{\baselineskip=10pt  \ General country and population distribution and statistics.}
\label{table: General country and population distribution and statistics}
\centering
\begin{tabular}{ll}
\hline
\hline
\multicolumn{1}{c}{Covariates (index $j$)} & \multicolumn{1}{c}{Explanation} \\ \hline
Total\_over\_65 (1) & Population ages 65 and above (\% of total population) in 2018.\\
Female\_per (2)& The percentage of female in the population in 2018.\\
Median\_age (5) &  Population median age in 2013.\\
Birth\_rate (6) & Crude birth rate (per 1000 population) in 2013.\\
Life\_expect\_total\_60 (14) & Life expectancy at age 60 (years) in 2016.\\
Dis\_to\_China (40) & Calculated by the R function \textsf{distm} based on the average \\
& longitude and latitude.\\
Popu\_density (44) & Population density (people per sq.km of land area) in 2018.\\
Tempe\_avg (45) & The average temperature in February and March in the captain\\
& of each country (we choose New York for US and Wuhan for \\
& China, due to the severe outbreak in the two cities).\\ \hline
\end{tabular}
\end{table}

\begin{table}[h]
\caption{\baselineskip=10pt  \ Health care resources.}
\label{table: Health care resources}
\centering
\begin{tabular}{ll}
\hline
\hline
\multicolumn{1}{c}{Covariates (index $j$)} & \multicolumn{1}{c}{Explanation} \\ \hline
Physician (3) & The number of physicians (per 1000 people) between \\
& 2015 and 2018.\\
Doc\_num\_per (12) & The number of medical doctors (per 10000 population)\\
& in 2016.\\
Hosp\_bed (13)&  Average hospital beds (per 10000 population) from \\
& 2013 to 2015.\\ \hline
\end{tabular}
\end{table}

\begin{table}[h]
\caption{\baselineskip=10pt  \ Tobacco and alcohol use.}
\label{table: Tobacco and alcohol use}
\centering
\begin{tabular}{ll}
\hline
\hline
\multicolumn{1}{c}{Covariates (index $j$)} & \multicolumn{1}{c}{Explanation} \\ \hline
Alcohol\_cons\_rec (7) & Recorded alcohol consumption per capita (15+) (in litres of \\
& pure alcohol), three-year average between 2015 and 2017.\\
Alcohol\_cons\_unrec (8) & Unrecorded alcohol consumption per capita (15+) (in litres \\
& of pure alcohol) in 2016.\\
Alcohol\_consumers\_total (9) & Alcohol consumers past 12 months (those adults who\\
& consumed alcohol in the past 12 months) (\% of total) in 2016.\\ 
Heavy\_drinking\_total (10) & Age-standardized estimates of the proportion of adults (15+ \\
& years) (who have had at least 60 grams or more of pure alcohol\\
& on at least one occasion in the past 30 days) in 2016.\\
Alcohol\_death\_total (11) & Alcohol-attributable death (\% of all-cause deaths in \\
& total) in 2016.\\
Tobacco\_smoke (34) & Age-standardized rates of prevalence estimates for daily \\
& smoking of any tobacco in adults (15+ years) in 2013.\\  
Cigarette\_smoke (35) & Age-standardized rates of prevalence estimates for daily \\
& smoking of any cigarette in adults (15+ years) in 2013.\\ \hline
\end{tabular}
\end{table}
 
\begin{table}[h]
\caption{\baselineskip=10pt  \ Disease and unhealthy prevalence.}
\label{table: Disease, unhealth, influenza virus detections}
\centering
\begin{tabular}{ll}
\hline
\hline
\multicolumn{1}{c}{Covariates (index $j$)} & \multicolumn{1}{c}{Explanation} \\ \hline
Underweight\_total (4) & Crude estimate of percent of adults with underweight \\
& (BMI $<$ 18.5) in 2016.\\
Blood\_glucose (30) & Age-standardized percent of 18+ population with raised fasting \\
& blood glucose ($\geq$7.0 mmol/L or on medication) in 2014.\\
Cholesterol (31) & Percentage of 25+ population with total cholesterol $\geq$ 240 mg/dl \\
& (6.2 mmol/l) in 2008.\\
Insuf\_phy\_act (32) & Age-standardized prevalence of insufficient physical activity \\
& (\% of adults aged 18+) in 2016. \\
Overweight (33) & Age-standardized prevalence of overweight among adults\\
&(BMI $\geq$ 25) (\% of adults aged 18+) in 2016.\\ 
Air\_pollution (36) & Concentrations of fine particulate matter (PM2.5) in 2016.\\
Air\_pollution\_death (37) & Age-standardized ambient air pollution attributable death rate\\
& (per 100000 population) in 2016.\\
Air\_pollution\_DALYs (38) & Age-standardized ambient air pollution attributable Disability-\\
& adjusted life year (DALYs) (per 100000 population) in 2016.\\
Tuberculosis\_case (39) & Incidence of tuberculosis (per 100000 population per year) in 2018.\\ \hline
\end{tabular}
\end{table}
 
\begin{table}[h]
\caption{\baselineskip=10pt  \ Testing and immunization statistics.}
\label{table: Testing, Immunization}
\centering
\begin{tabular}{ll}
\hline
\hline
\multicolumn{1}{c}{Covariates (index $j$)} & \multicolumn{1}{c}{Explanation} \\ \hline
Diphtheria tetanus toxoid and pertussis & Diphtheria tetanus toxoid and pertussis third-dose  \\
 third-dose immunization (15)& (DTP3) immunization coverage (\% of total \\
 &  1-year-olds) in 2018. \\
Hepatitis B third-dose   & Hepatitis B third-dose (HepB3) immunization coverage \\
immunization (16) & (\% of total 1-year-olds) in 2018.\\
Haemophilus influenzae type B  & Haemophilus influenzae type B third-dose (Hib3)   \\
 third-dose immunization (17)& immunization coverage (\% of total 1-year-olds) in 2018.\\
Measles-containing-vaccine & Measles-containing-vaccine first-dose (MCV1)  \\
first-dose immunization (18) & immunization coverage (\% of total 1-year-olds) \\
 &  in 2018.\\
Measles-containing-vaccine   & Measles-containing-vaccine second-dose (MCV2)  \\
second-dose immunization (19)& immunization coverage (\% of total nationally \\
 & recommended age) in 2018. \\
Pneumococcal conjugate vaccines   & Pneumococcal conjugate vaccines third-dose (PCV3) \\
third-dose immunization (20)& immunization coverage (\% of total 1-year-olds) in 2018.\\
Polio third-dose immunization (21) & Polio (Pol3) third-dose immunization coverage  \\
& (\% of total 1-year-olds) in 2018.\\
Testing\_num (41) & The number of COVID-19 testing cases\\ &(\href{https://ourworldindata.org/covid-testing}{ourworldindata.org/-} collect the data and the data dates \\
& are between Febrary and March on several media).\\
Testing\_confirm (42) & The total number of confirmed cases on \\
& the same day with testing\_num divided by the \\
& covariate Testing\_num\_COVID19 (41).\\
Testing\_popu (43) & The covariate Testing\_num\_COVID19 (41) divided \\
& by covariate Total\_popu (2).\\ \hline
\end{tabular}
\end{table}
 
\begin{table}[h]
\caption{\baselineskip=10pt  \ International health regulations (IHR) monitoring framework.}
\label{table: International Health Regulations monitoring framework1}
\centering
\begin{tabular}{ll}
\hline
\hline
\multicolumn{1}{c}{Covariates (index $j$)} & \multicolumn{1}{c}{Explanation} \\ \hline
Zoonotic\_Events (22) & Scores that show whether mechanisms for detecting \\
& and responding to zoonoses and potential zoonoses are \\
& established and functional in 2018. \\
Food\_Safety (23) & Scores that show whether mechanisms are established \\
& and functioning for detecting and responding to \\
& foodborne disease and food contamination in 2018.\\
Laboratory (24) & Scores that show the availability of laboratory \\
& diagnostic and confirmation services to test for priority\\
& health threats in 2018.\\
Human\_Resources (25) & Scores that show the availability of human resources \\
& to implement IHR Core Capacity. \\
Health\_Service\_Provision (26) & Scores that show an immediate output of the inputs \\
& into the health system, such as the health workforce, \\
& procurement and supplies, and financing in 2018.\\ 
Risk\_Communication (27) & Scores that show mechanisms for effective risk \\
& communication during a public health emergency \\
& are established and functioning in 2018.\\
Points\_of\_Entry (28) & Scores that show whether general obligations\\
& at point of entry are fulfilled (including for \\
& coordination and communication) to prevent the \\
& spread of diseases through international traffic in 2018.\\
Radiation\_Emergencies (29) & Scores that show whether mechanisms are established \\
& and functioning for detecting and responding to \\
& radiological and nuclear emergencies that may constitute\\
& a public health event of international concern in 2018.\\ \hline
\end{tabular}
\begin{itemize}
    \item[] NOTE 1: The International health regulations, or IHR (2005), represent an agreement between 196 countries including all WHO Member States to work together for global health security. Through IHR, countries have agreed to build their capacities to detect, assess, and report public health events. WHO plays the coordinating role in IHR and, together with its partners, helps countries to build capacities. (\href{https://www.who.int/ihr/about/en/}{https://www.who.int/ihr/about/-})
    \item[] NOTE 2: IHR monitoring framework was developed, which represents a consensus among technical experts from WHO Member States, technical institutions, partners and WHO. (\href{https://www.who.int/ihr/procedures/monitoring/en/}{https://www.who.int/ihr/procedures/-})
\end{itemize}
\end{table}

\clearpage

% Include only the SI item label in the paragraph heading. Use the \nameref{label} command to cite SI items in the text.

\section*{S2 Appendix: Technical expressions for the three models $\mathcal{M}_1$, $\mathcal{M}_2$, and $\mathcal{M}_3$}\label{S2_Appendix}

Technical expressions for the three models, $\mathcal{M}_1$, $\mathcal{M}_2$, and $\mathcal{M}_3$, compared in the main paper are given as follows:
\begin{itemize}
\baselineskip=15pt
    \item[] $\mathcal{M}_1$ is an individual country-based model (nonhierarchical model) that uses infection trajectory for a single country $\textbf{y} = (y_{1},\cdots,y_{T})^{\top}$. The model is given by
    \begin{align*}
y_{t}&=f(t; \theta_{1},\theta_{2},\theta_{3}, \xi) + \epsilon_{t},\, \epsilon_{t}\sim \mathcal{N}(0,\sigma^2),\,
\theta_{l}  \sim \mathcal{N}(\alpha_l, \sigma_l^2 ),\,\\ \xi&\sim \log\ \mathcal{N}(0,1),
 (t = 1, \cdots, T,\, l = 1,2,3),
\end{align*}
where $f(t;\theta_{1},\theta_{2},\theta_{3})$ is the Richards growth curve (\ref{eq:Richards growth curve}), and improper priors \citep{gelman2004bayesian} are used for error variances and intercept terms as (\ref{eq:improper_priors}).
    \item[] $\mathcal{M}_2$ is a Bayesian hierarchical model without using covariates, which uses infection trajectories from $N$ countries, $\{\textbf{y}_i\}_{i=1}^{N}$. This model is equivalent to BHRM (\ref{eq:likelihood_part}) -- (\ref{eq:improper_priors}) with removed covariates terms in (\ref{eq:linear_regression}).
    \item[] $\mathcal{M}_3$ is the BHRM (\ref{eq:likelihood_part}) -- (\ref{eq:improper_priors}).
\end{itemize}

\section*{S3 Appendix: Posterior computation}

We illustrate a full description of a posterior computation for the BHRM (\ref{eq:likelihood_part}) -- (\ref{eq:improper_priors}) by using a Markov chain Monte Carlo (MCMC) simulation \citep{robert2013monte}. To start with, for illustrative purpose, we shall use vectorized notations for the likelihood part (\ref{eq:likelihood_part}), regression part (\ref{eq:linear_regression}), and its coefficients part (\ref{eq:Horseshoe}):
\begin{align*}
\textbf{y}_{i}|\theta_{1i},\theta_{2i},\theta_{3i}, \xi_i,\sigma^{2}
&\sim
\mathcal{N}_{T}(
\textbf{f}( \theta_{1i},\theta_{2i},\theta_{3i}, \xi_i) 
,\sigma^{2} \mathbf{I}),\quad (i = 1,\cdots,N),
\\
\bm{\theta}_{l} | \alpha_{l}, \bm{\beta}_{l}, \sigma_{l}^{2}
&\sim
\mathcal{N}_{N}( 
\textbf{1} \alpha_{l} + \textbf{X} \bm{\beta}_{l}
,\sigma_{l}^{2} \mathbf{I})
,\quad\quad\quad\quad (l = 1,2,3),\\
\bm{\beta}_{l}|\tau_l, \bm{\lambda}_l, \sigma_l^2
&\sim
\mathcal{N}_{p}(
\textbf{0},
\sigma_{l}^{2}\tau_{l}^{2}
\bm{\Lambda}_{l})
,\quad\quad\quad\quad\quad\quad (l = 1,2,3).
\end{align*}
The $T$-dimensional vector $\textbf{y}_i = (y_{i1},\cdots,y_{it}, \cdots, y_{iT})^{\top}$ ($i=1,\cdots,N$) is the observed infection trajectory for the country $i$ across the times. The notation $\textbf{f}(\theta_{1i},\theta_{2i},\theta_{3i},\xi_i)$ ($i=1,\cdots,N$) is $T$-dimensional vector that describes the Richard curves across the times:
\begin{align*}
\textbf{f}(\theta_{1i},\theta_{2i},\theta_{3i},\xi_i)
=
(
f(1 ; \theta_{1i}, \theta_{2i}, \theta_{3i},\xi_i)
,
\dots,
f(T ;\theta_{1i}, \theta_{2i}, \theta_{3i},\xi_i)
)^{\top},\quad (i = 1,\cdots,N).
\end{align*}
The vectors $\bm{\theta}_{l}=(\theta_{l1},\cdots,\theta_{lN})^{\top}$ ($l=1,2,3$) and $\bm{\xi} = (\xi_{1},\cdots,\xi_{N})^{\top}$ are $N$-dimensional vectors for the four parameters of the Richards curve (\ref{eq:Richards growth curve}) across the $N$ countries.

The matrix $\textbf{X}$ is $N$-by-$p$ design matrix whose $i$-th row vector is given by the $p$ predictors $\textbf{x}_{i}=
(x_{i1},\cdots, x_{ip})^{\top} \in \mathbb{R}^p $, $(i = 1, \cdots, N)$. The notation $\mathbf{I}$ stands for an identity matrix. Before implementing, it is recommended that
each of column vectors of the design matrix $\textbf{X}$ is standardized \citep{tibshirani1996regression,armagan2013generalized}: that is, each column vector has been centered, and then columnwisely scaled so that each column vector has mean zero and unit Euclidean norm ($l_2$-norm).

The $p$-dimensional vector $\bm{\beta}_{l} = (\beta_{l1},\cdots,\beta_{lp})^{\top}$ ($l=1,2,3$) denotes $p$ coefficients from the $l$-th regression. The vector $\bm{\lambda}_l = (\lambda_{l1}, \cdots,\lambda_{lp})^{\top}$ ($l=1,2,3$) is $p$-dimensional vector for the local-scale parameters, and the matrix $\bm{\Lambda}_{l}$ is $p$-by-$p$ diagonal matrix $\bm{\Lambda}_{l} = \text{diag}(\lambda_{l1}^2,\cdots, \lambda_{lp}^2)$ ($l=1,2,3$). The $\tau_l$ ($l=1,2,3$) is referred to as the global-scale parameter \citep{carvalho2010}.

Under the formulation of BHRM (\ref{eq:likelihood_part}) -- (\ref{eq:improper_priors}), our goal is to sample from the full joint posterior distribution $\pi(\bm{\theta}_{1}, \bm{\theta}_{2}, \bm{\theta}_{3},\bm{\xi},\sigma^2, \Omega_1,\Omega_2,\Omega_3|\textbf{y}_{1:N})$ where $\Omega_{l} = \{\alpha_{l}, \bm{\beta}_{l},\bm{\lambda}_{l}, \tau_l, \sigma_{l}^{2} \}$ ($l = 1,2,3$), whose proportional part is given by
\begin{align*}
\bigg\{
\prod_{i =1}^{N}
 \mathcal{N}_{T}(\textbf{y}_{i}|
\textbf{f}_{i}( \theta_{1i},\theta_{2i},\theta_{3i}, \xi_i) 
,\sigma^{2} \mathbf{I})
\bigg\}
&\bigg\{
\prod_{l =1}^{3}
\mathcal{N}_{N}
(
\bm{\theta}_{l}
|
\textbf{1}
\alpha_{l} + \textbf{X}\bm{\beta}_{l},\sigma_{l}^2 \mathbf{I})
\mathcal{N}_{p}(\bm{\beta}_{l}
| 
\textbf{0},
\sigma_{l}^{2}\tau_{l}^{2}
\bm{\Lambda}_{l})
\pi( \bm{\lambda}_{l})
\pi( \tau_{l})
\pi( \sigma_{l}^{2})
\bigg\}\\
&\cdot
\bigg\{
\prod_{i =1}^{N}
\log\ \mathcal{N}(\xi_i|0,1)
\bigg\}
\pi(\sigma).
%%%%%%%%%%%%%%%%%%%%%%%%%%%%%%%%%%%%%%%%%%%%%%%%%%%%%%%%%%%%
\end{align*}
\noindent
To sample from the full joint density, we use a Gibbs sampler \citep{casella1992explaining} to exploit conditional independences among the latent variables induced by the hierarchy. The following algorithm describes a straightforward Gibbs sampler
\begin{itemize}
\baselineskip=15pt
\item[] \textbf{\emph{Step 1.}} 
Sample $\bm{\theta}_{1}$ from its full conditional distribution
\begin{align*}
\pi(
\bm{\theta}_{1} 
|
-)
&\sim 
\mathcal{N}_{N}
(
\bm{\Sigma}_{\bm{\theta}_{1}}
\{
(1/\sigma^{2})
\textbf{r}
+ 
(1/\sigma_l^{2})
(\textbf{1}\alpha_1 + \textbf{X}\bm{\beta}_1)
\}
,
\bm{\Sigma}_{\bm{\theta}_{1}}
), 
\end{align*}
where
$
\bm{\Sigma}_{\bm{\theta}_{1}}
=
\{
(1/\sigma^{2})
\textbf{H}
+
(1/\sigma_{l}^{2})
\textbf{I}
\}^{-1}
\in \mathbb{R}^{N \times N}
$. The matrix $\textbf{H}$ is $N$-by-$N$ diagonal matrix $\textbf{H}=\text{diag}(\|\textbf{h}(\theta_{21},\theta_{31},\xi_1)\|_2^2, \cdots, \|\textbf{h}(\theta_{2N},\theta_{3N},\xi_N)\|_2^2)$, and the vector \textbf{r} is a $N$-dimensional vector which is given by
$\textbf{r} =
(
\textbf{y}_{1}^{\top}
\textbf{h}(\theta_{21},\theta_{31},\xi_1)
,
\dots
,
\textbf{y}_{N}^{\top}
\textbf{h}(\theta_{2N},\theta_{3N},\xi_N)
)^{\top}$, where each of the $T$-dimensional vector $\textbf{h}(\theta_{2i},\theta_{3i},\xi_i)$ ($i=1,\cdots, N$) is obtained by
\begin{align*}
\textbf{h}(\theta_{2i}, \theta_{3i},\xi_i)
=
(
h(1 ; \theta_{2i}, \theta_{3i},\xi_i)
,
\dots,
h(T ; \theta_{2i}, \theta_{3i}, \xi_i)
)^{\top},
\end{align*}
where $h(t ; \theta_2, \theta_3,\xi) = [ 1 + \xi \cdot \exp \{-\theta_2 \cdot ( t - \theta_3) \}   ]^{-1/\xi}$.
\item[]\textbf{\emph{Step 2.}} Sample $\theta_{2i}$ and $\theta_{3i}$, $i=1,\cdots, N$, independently from their full conditional distributions. Proportional parts of the distributions are given by
\begin{align}
%\label{eq: proportional parts of theta_2i}
\nonumber
\pi(
\theta_{2i}
|
-)
 & \propto 
 \exp
\bigg(
-\frac{1}{2 \sigma^{2}}
\|
\textbf{y}_{i}
-
\textbf{f}(\theta_{1i}, \theta_{2i}, \theta_{3i},\xi_i)
\|_2^{2}
-\frac{1}{ 2 \sigma_{2}^{2}}
(\theta_{2i} - \alpha_{2} - \textbf{x}_{i}^{\top} \bm{\beta}_{2})^{2}
\bigg) ,
 \\
 %\label{eq: proportional parts of theta_3i}
\nonumber
\pi(
 \theta_{3i}
|
-)
 &
 \propto
\exp
\bigg(
-\frac{1}{2 \sigma^{2}}
\|
\textbf{y}_{i}
-
\textbf{f}(\theta_{1i}, \theta_{2i}, \theta_{3i},\xi_i)
\|_2^{2}
-\frac{1}{ 2 \sigma_{3}^{2}}
(\theta_{3i} - \alpha_{3} - \textbf{x}_{i}^{\top} \bm{\beta}_{3})^{2}
\bigg).
\end{align}
Here, $\| \cdot \|_2$ indicates the $l_2$-norm. Note that the two conditional densities are not known in closed forms because two parameters, $\theta_{2i}$ and $\theta_{3i}$, participate to the function $\textbf{f}(\theta_{1i}, \theta_{2i}, \theta_{3i},\xi_i)$ in a nonlinear way. We use the Metropolis algorithm \citep{andrieu2003introduction} with Gaussian proposal densities within this Gibbs sampler algorithm. 
\item[] \textbf{\emph{Step 3.}} Sample $\xi_{i}$, $i=1,\cdots, N$, independently from its full conditional distribution. Proportional parts of the distributions are given by
\begin{align}
\label{eq:full_condi_xi}
    \pi(\xi_i|-) &\propto
     \exp
\bigg(
-\frac{1}{2 \sigma^{2}}
\|
\textbf{y}_{i}
-
\textbf{f}(\theta_{1i}, \theta_{2i}, \theta_{3i},\xi_i)
\|_2^{2}
\bigg)
\cdot
\log\ \mathcal{N}(\xi_i|0,1).
\end{align}
Note that the density (\ref{eq:full_condi_xi}) is not expressed in a closed form distribution. Because the shape parameter $\xi_i$ is supported on $(0,\infty)$ and participates in the Richards curve (\ref{eq:Richards growth curve}) as an exponent, sampling from the density needs a delicate care, where by we employed the elliptical slice sampler \citep{murray2010elliptical}.
\item[] \textbf{\emph{Step 4.}} 
Sample $\sigma^{2}$ from its full conditional distribution
\begin{align*}
\pi
(
\sigma^{2}
|
-)
&
\sim
\mathcal{IG}
\bigg(
\frac{NT}{2}
, 
\frac{1}{2}
\sum_{i=1}^{N}
\|
\textbf{y}_{i}
- 
\textbf{f}
(\theta_{1i}, \theta_{2i}, \theta_{3i},\xi_i )
\|_2^{2}
\bigg).
\end{align*}
\item[] \textbf{\emph{Step 5.}} Sample $\alpha_{l}$, $l =1,2,3$, independently from their full conditional distributions
\begin{align*}
\pi
(
\alpha_{l} |
-
)
&\sim
\mathcal{N}_{1}( \bm{1}^{\top} (\bm{\theta}_{l} -  \textbf{X} \bm{\beta}_{l} )/N 
, 
\sigma_{l}^{2}/N
 ).
\end{align*}
   \item[]\textbf{\emph{Step 6}.} Sample $\bm{\beta}_{l}$, $l =1,2,3$, independently from conditionally independent posteriors
\begin{align*}
\pi
(
\bm{\beta}_{l} 
|
-)
&\sim
\mathcal{N}_{p}
(\bm{\Sigma}_{\bm{\beta}_{l}} \textbf{X}^{\top} (\bm{\theta}_{l} - \bm{1}\alpha_{l})
, \sigma_{l}^{2} \bm{\Sigma}_{\bm{\beta}_{l}} ),
\end{align*}
where $ \bm{\Sigma}_{\bm{\beta}_{l}} = [\textbf{X}^{\top} \textbf{X} + \Lambda_{*l}^{-1}]^{-1
}
\in \mathbb{R}^{p \times p}$, 
$\bm{\Lambda}_{l} = \text{diag}(\lambda_{l1}^{2}, \cdots, \lambda_{lp}^{2}) \in \mathbb{R}^{p \times p}$, and $
\bm{\Lambda}_{*l} = \tau_l^{2} \bm{\Lambda}_{l}$.
\item[] \textbf{\emph{Step 7.}} Sample $\lambda_{lj}$, $l=1,2,3,\,j=1,\cdots,p$, independently from conditionally independent posteriors
\begin{align*}
\pi
(
\lambda_{lj}
|
-)
&\sim
\mathcal{N}(\beta_{lj}|
0,
\sigma_l^2
\tau_l^{2} 
\lambda_{lj}^{2} 
)
\cdot
\{1/(1+\lambda_{lj}^2) \}
.
\end{align*}
Note that the densities $\pi
(
\lambda_{lj}
|
-)$ $(l=1,2,3, \,j=1,\cdots,p)$ are not expressed in closed forms: we use the slice sampler \citep{neal2003slice}.
\item[] \textbf{\emph{Step 8.}} Sample $\tau_{l}$, $l=1,2,3$, independently from conditionally independent posteriors
\begin{align*}
\pi
(
\tau_{l}
|
-)
&\sim
\mathcal{N}_{p}(\bm{\beta}_{l}|
\textbf{0},
\sigma_l^2
\tau_l^{2}
\bm{\Lambda}_{l}
)
\cdot
\{1/(1+\tau_l^2) \}
.
\end{align*}
Note that the densities $\pi
(
\tau_{l}
|
-)$ $(l=1,2,3)$ are not expressed in closed forms: we use the slice sampler \citep{neal2003slice}.
\item[] \textbf{\emph{Step 9.}} Sample $\sigma_{l}^{2}$, $l =1,2,3$, independently from their full conditionally distributions
\begin{align*}
%\label{eq: sampling sigma in non-spatial}
 \pi(\sigma_{l}^{2}
|-)
 &\sim 
\mathcal{IG}
\bigg(\frac{N  + p}{2}, 
\frac{\| \bm{\theta}_{l} - \bm{1}\alpha_{l} - \textbf{X}  \bm{\beta}_{l} \|_2^{2} +  
\bm{\beta}_{l}^{\top} \bm{\Lambda}_{*l}^{-1}\bm{\beta}_{l}}{2}
 \bigg).
\end{align*}
\end{itemize}
\subsection*{Elliptical slice sampler for Step 3}\label{subsec:Elliptical slice sampler}
To start with we shall use the variable change ($\eta = \log \xi$) to the right hand side of (\ref{eq:full_condi_xi}):
\begin{align}
\label{eq:full_condi_eta}
    \pi(\eta_i|-) &\propto
    \mathcal{L}(\eta_i)
\cdot
\mathcal{N}(\eta_i|0,1),\,\quad(i = 1,\cdots,N),
\end{align}
such that $\mathcal{L}(\eta_i)
=
\exp
\{
-
\|
\textbf{y}_i
-
\textbf{f}(\theta_{1i}, \theta_{2i}, \theta_{3i}, e^{\eta_i} )
\|_2^{2}
/(2 \sigma^{2})
\}$ corresponds to a likelihood part.

Now, we use the elliptical slice sampler (ESS) \citep{murray2010elliptical,nishihara2014parallel} to sample from $\eta_i^{(s+1)} \sim \pi(\eta_i|-)$ (\ref{eq:full_condi_eta}) ($i=1,\cdots,N$) that exploits the Gaussian prior measure. Conceptually, ESS and the Metropolis-Hastings (MH) algorithm \citep{chib1995understanding} are similar: both methods are comprised of two steps: \emph{proposal step} and \emph{criterion step}. A difference between the two algorithms arises in the \emph{criterion step}. If the new candidate does not pass the criterion, then MH takes the current state as the next state: whereas, ESS re-proposes a new candidate until rejection does not take place, rendering the algorithm rejection-free. Further information for ESS is referred to the original paper \citep{murray2010elliptical}. After ESS has been employed, the realized $\eta_i$ needs to be transformed back through $\xi = e^\eta$. Algorithm \ref{alg:Elliptical slice sampler centered by the Hill estimator} illustrates the full description of algorithms: to avoid notation clutter, the index $i$ is omitted.
\begin{algorithm}\label{eq:ESS_xi}
\baselineskip=10pt
 \caption{Elliptical slice sampler to sample from $\pi(\xi|-)$}\label{alg:Elliptical slice sampler centered by the Hill estimator}
\SetAlgoLined
\textbf{Circumstance : } At the \textbf{\emph{Step 3}} of the $s$-th iteration of the Gibbs sampler.\\
\textbf{Input : } Current state $\xi^{(s)}$.\\
\textbf{Output : } A new state $\xi^{(s+1)}$.\\
1. Variable change ($\eta = \log \xi$): $\eta^{(s)} = \log\ \xi^{(s)}$.\\
2. Implement elliptical slice sampler;\\
\begin{itemize}
\item[a. ] Choose an ellipse : $\nu \sim \mathcal{N}_{1}(0,1)$.
\item[b. ] Define a criterion function: 
\begin{align*}
\alpha(\eta, \eta^{(s)}) = \text{min}\{\mathcal{L}(\eta) / \mathcal{L}(\eta^{(s)}),1\}: \mathbb{R} \rightarrow [0,1],
\end{align*}
where $\mathcal{L(\eta)}
=
\exp
\{
-
\|
\textbf{y}
-
\textbf{f}(\theta_{1}, \theta_{2}, \theta_{3}, e^{\eta} )
\|_2^{2}
/(2 \sigma^{2})
\}
$.
\item[c. ] Choose a threshold and fix: $u \sim \mathcal{U}[0,1]$.
\item[d. ] Draw an initial proposal $\eta^{*}$: 
\begin{align*}
\phi &\sim \mathcal{U}(-\pi,\pi]\\
\eta^{*}&=\eta^{(s)} \cos\ \phi
+
\nu \sin\ \phi
\end{align*}
\item[e. ] \textbf{if}  \textbf{(} $u < \alpha(\eta^{*}, \eta^{(s)})$  \textbf{)} $\{$ $\eta^{(s+1)} = \eta^{*}$  $\}$ \textbf{else} $\{$ \\
\indent$\quad$ Define a bracket : $(\phi_{\text{min}},\phi_{\text{max}} ]= (-\pi, \pi ]$.\\
\indent$\quad$ \textbf{while} \textbf{(} $u \geq \alpha(\eta^{*}, \eta^{(s)})$ \textbf{)} $\{$\\
\indent$\quad\quad$ Shrink the bracket and try a new point :
\\
\indent$\quad\quad$ \textbf{if} \textbf{(} $\phi > 0$ \textbf{)} $\phi_{\text{max}} = \phi$ \textbf{else} $\phi_{\text{min}} = \phi$ \\
\indent$\quad\quad\quad$ $\phi \sim \mathcal{U}(\phi_{\text{min}},\phi_{\text{max}}]$\\
\indent$\quad\quad\quad$
$\eta^{*}=\eta^{(s)} \cos\ \phi
+
\nu \sin\ \phi$
\\
\indent$\quad\quad$ $\}$\\
\indent$\quad\quad$ $\eta^{(s+1)} = \eta^{*}$\\
\indent$\quad$ $\}$
\end{itemize}
4. Variable change ($\xi = e^{\eta}$): $\xi^{(s+1)} = \exp\ \eta^{(s+1)}$.\\
\end{algorithm}

\clearpage

\section*{S4 Appendix: Infection trajectories for the top 20 countries}

The section includes extrapolated infection trajectories for the top 20 countries that are most severely affected by the COVID-19.

\begin{figure}[H]
    \centering
    \includegraphics[scale = 0.42]{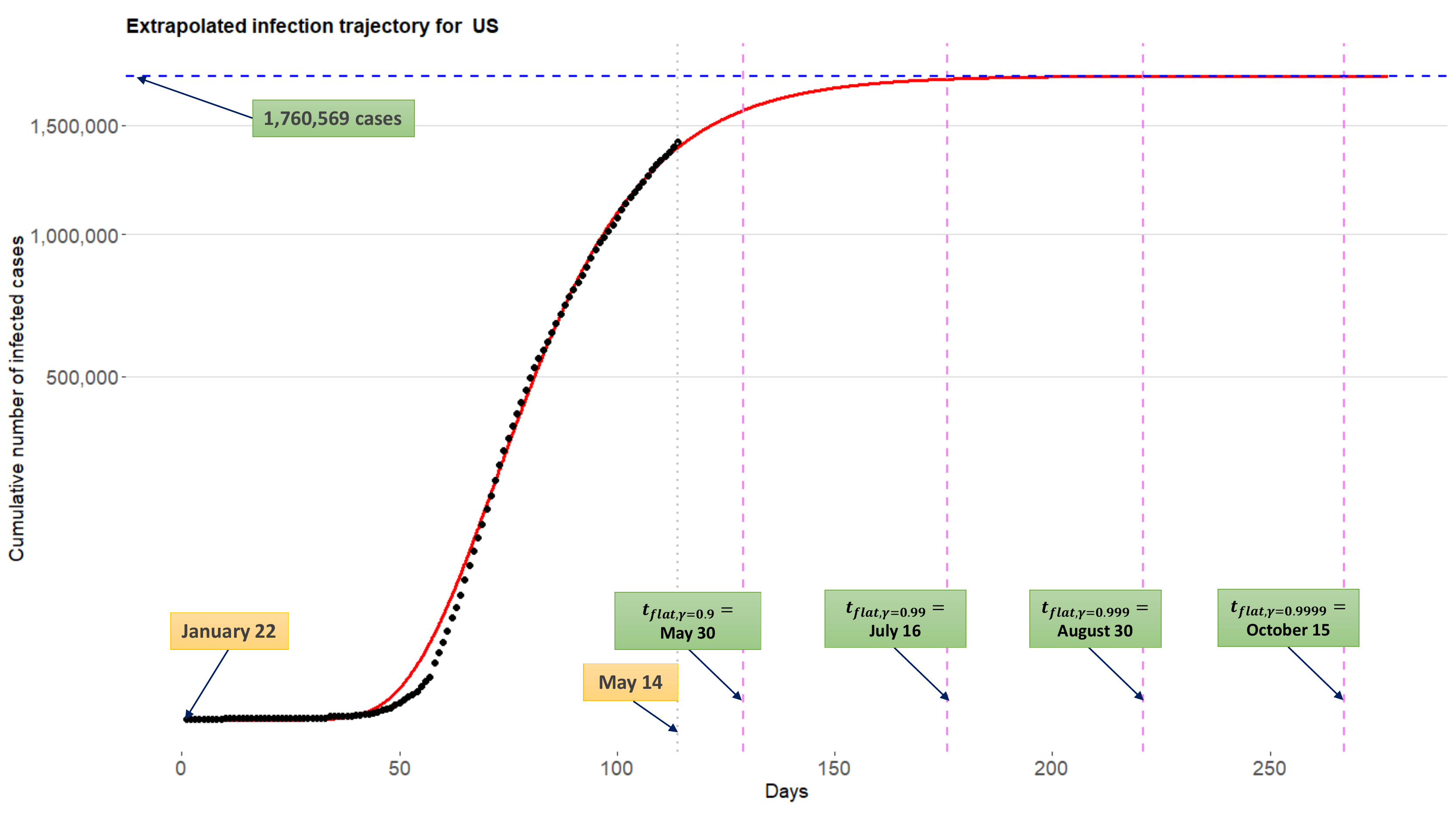}
    \caption{\baselineskip=10pt   Extrapolated infection trajectory for the US based on the model $\mathcal{M}_3$.}
    \label{fig:US}
\end{figure}

\begin{figure}[H]
    \centering
    \includegraphics[scale = 0.42]{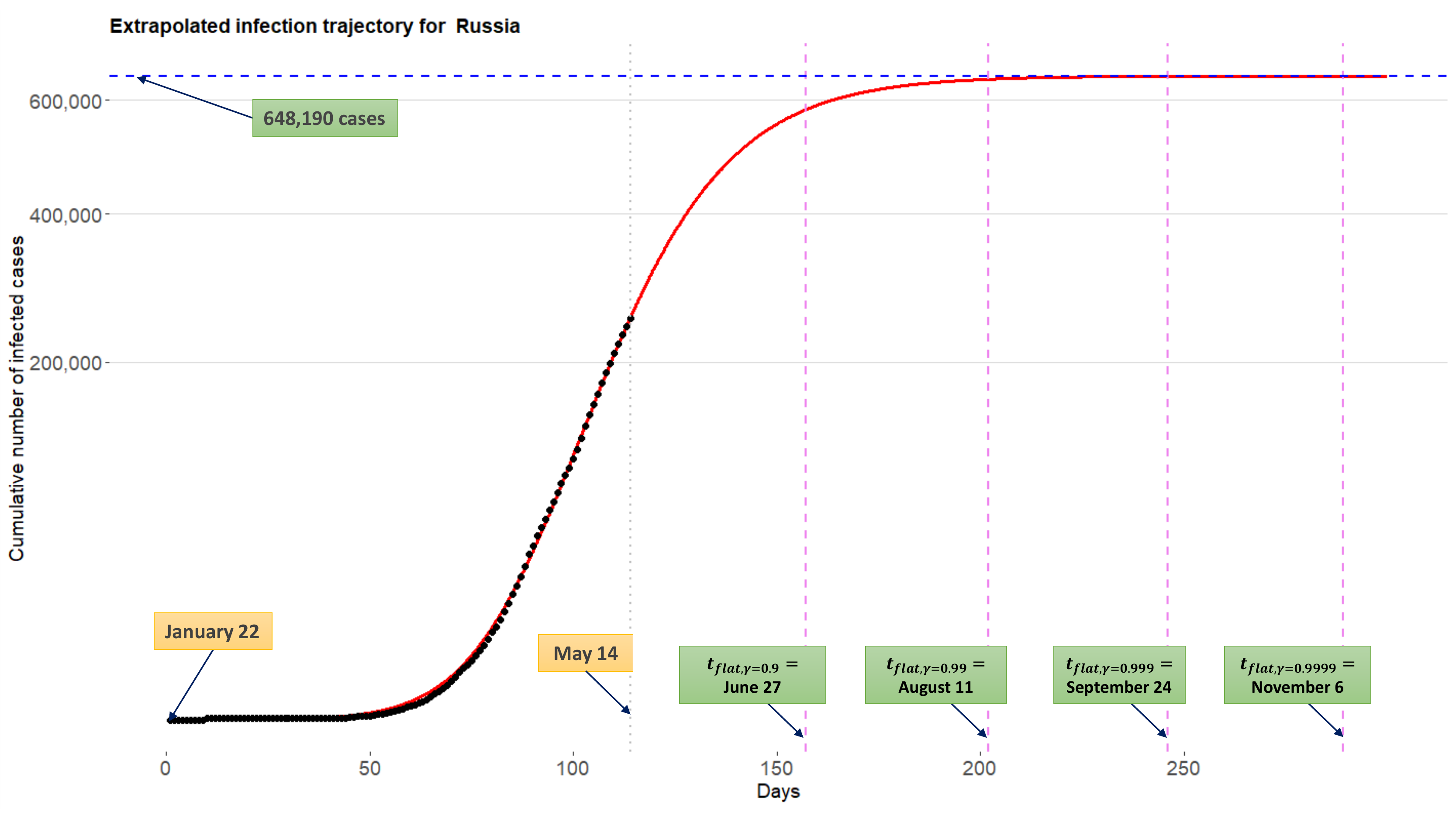}
    \caption{\baselineskip=10pt   Extrapolated infection trajectory for the Russia based on the model $\mathcal{M}_3$.}
    \label{fig:Russia}
\end{figure}

\begin{figure}[H]
    \centering
    \includegraphics[scale = 0.42]{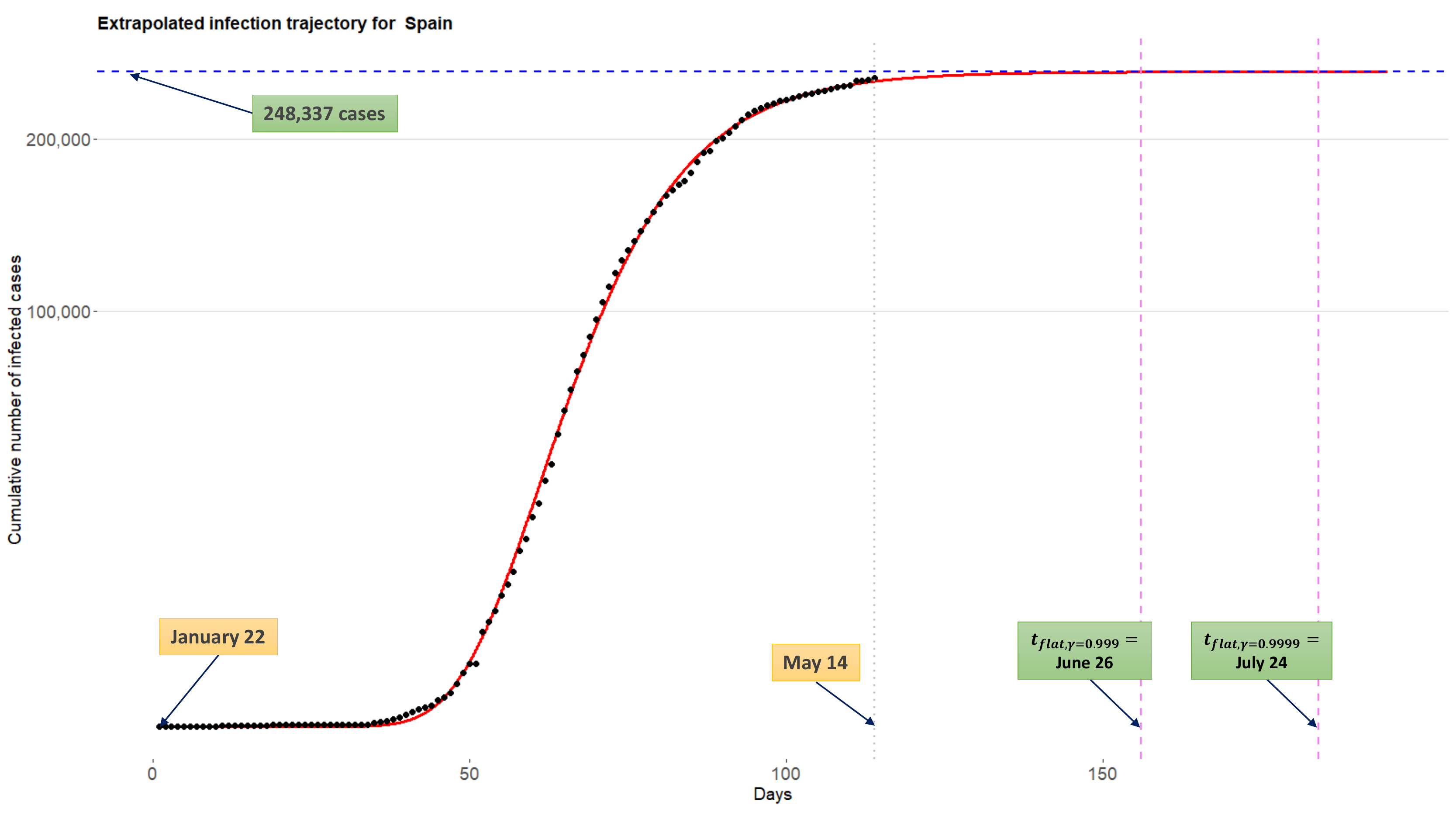}
    \caption{\baselineskip=10pt   Extrapolated infection trajectory for the Spain based on the model $\mathcal{M}_3$.}
    \label{fig:Spain}
\end{figure}

\begin{figure}[H]
    \centering
    \includegraphics[scale = 0.42]{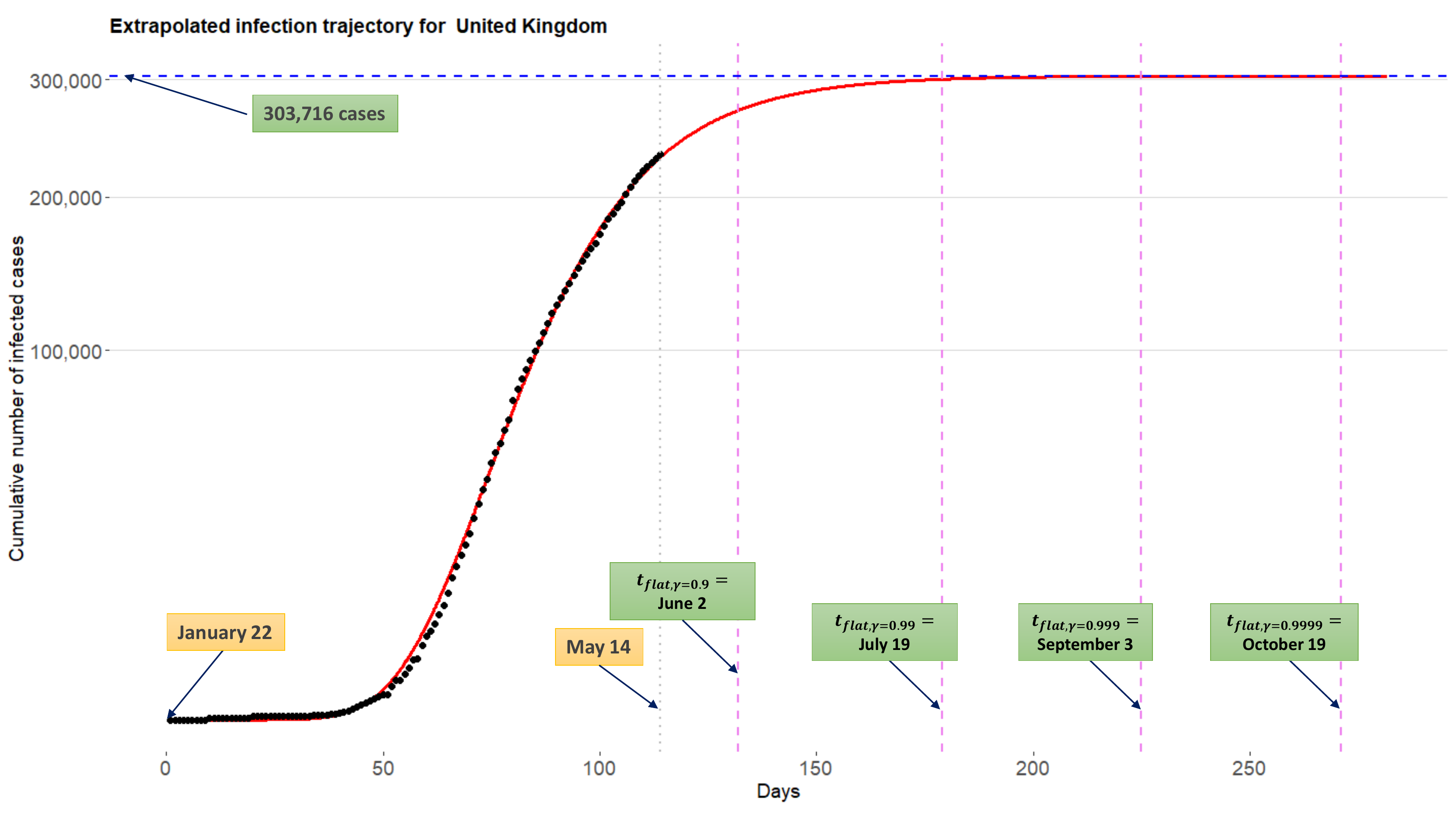}
    \caption{\baselineskip=10pt   Extrapolated infection trajectory for the United Kingdom based on the model $\mathcal{M}_3$.}
    \label{fig:UK}
\end{figure}

\begin{figure}[H]
    \centering
    \includegraphics[scale = 0.42]{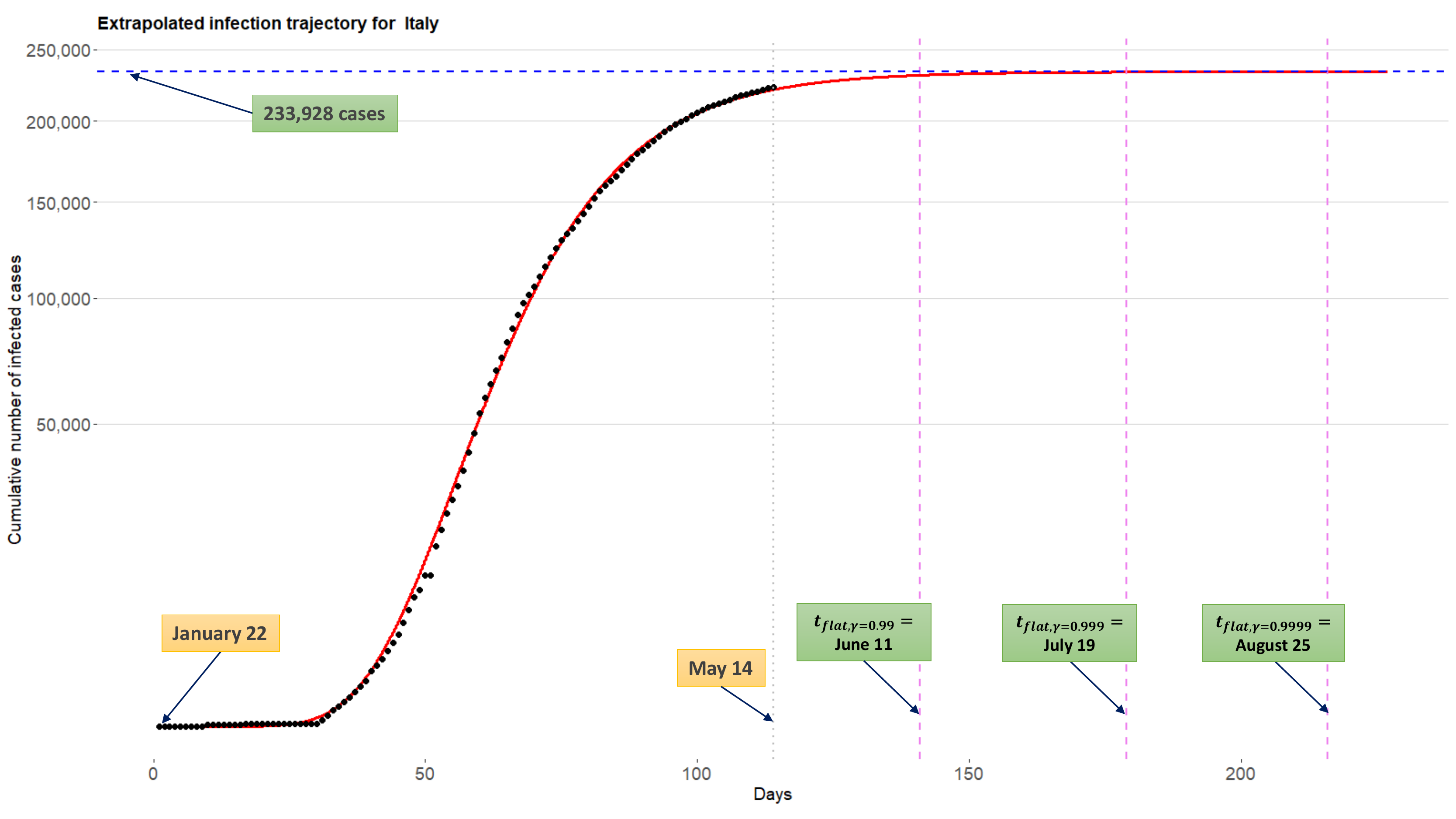}
    \caption{\baselineskip=10pt   Extrapolated infection trajectory for the Italy based on the model $\mathcal{M}_3$.}
    \label{fig:Italy}
\end{figure}

\begin{figure}[H]
    \centering
    \includegraphics[scale = 0.42]{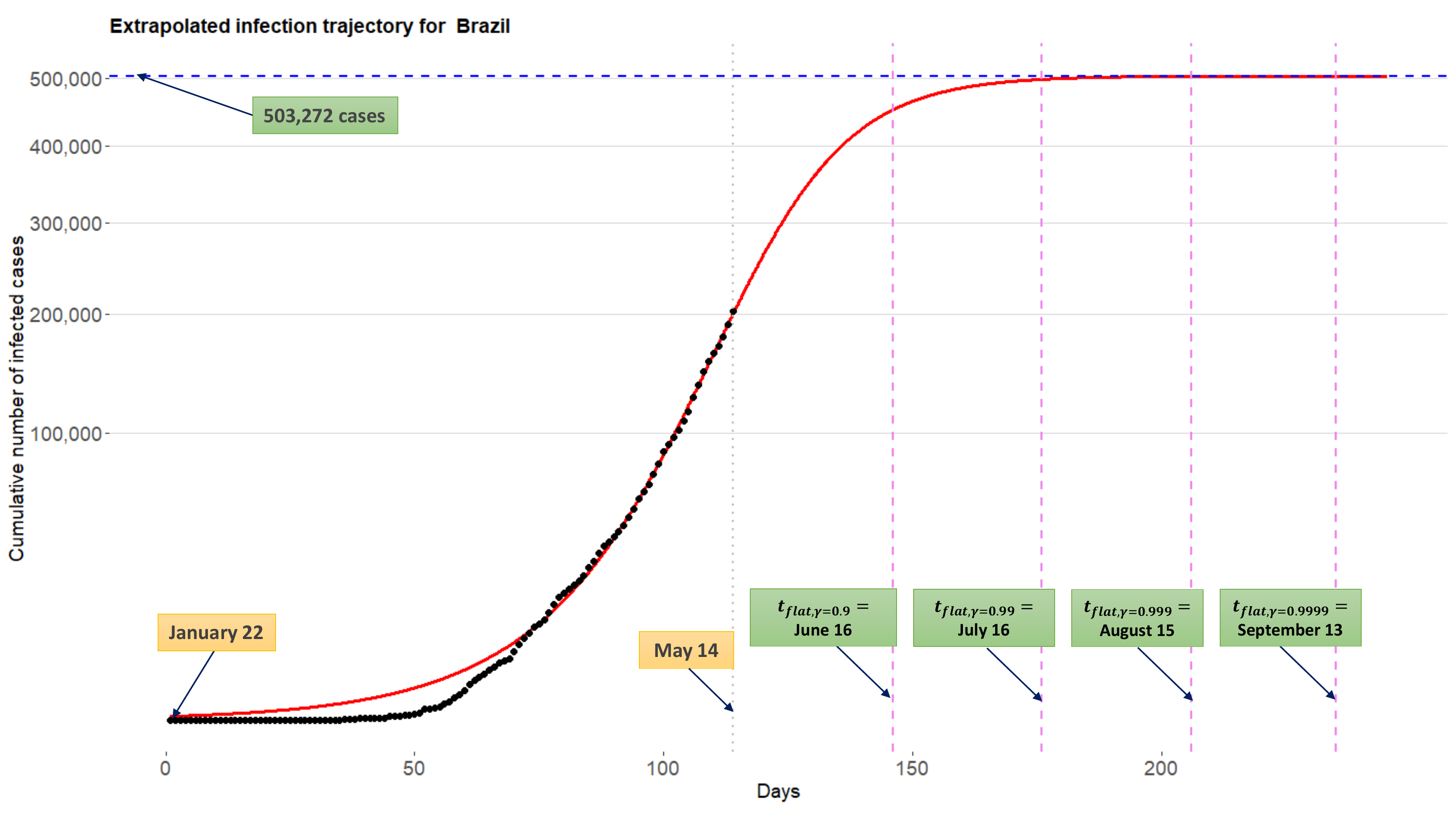}
    \caption{\baselineskip=10pt   Extrapolated infection trajectory for the Brazil based on the model $\mathcal{M}_3$.}
    \label{fig:Brazil}
\end{figure}

\begin{figure}[H]
    \centering
    \includegraphics[scale = 0.42]{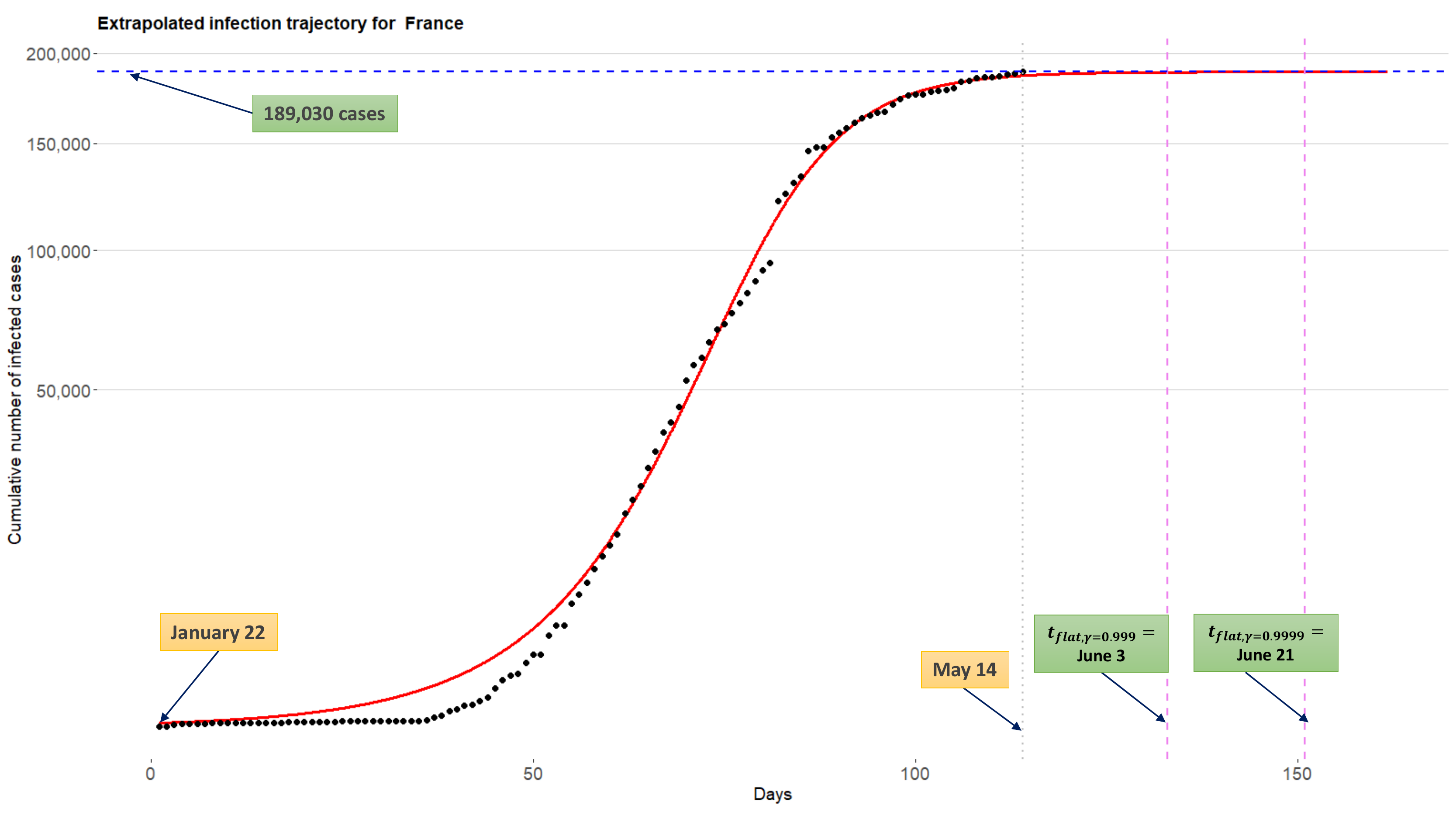}
    \caption{\baselineskip=10pt   Extrapolated infection trajectory for the France based on the model $\mathcal{M}_3$.}
    \label{fig:France}
\end{figure}

\begin{figure}[H]
    \centering
    \includegraphics[scale = 0.42]{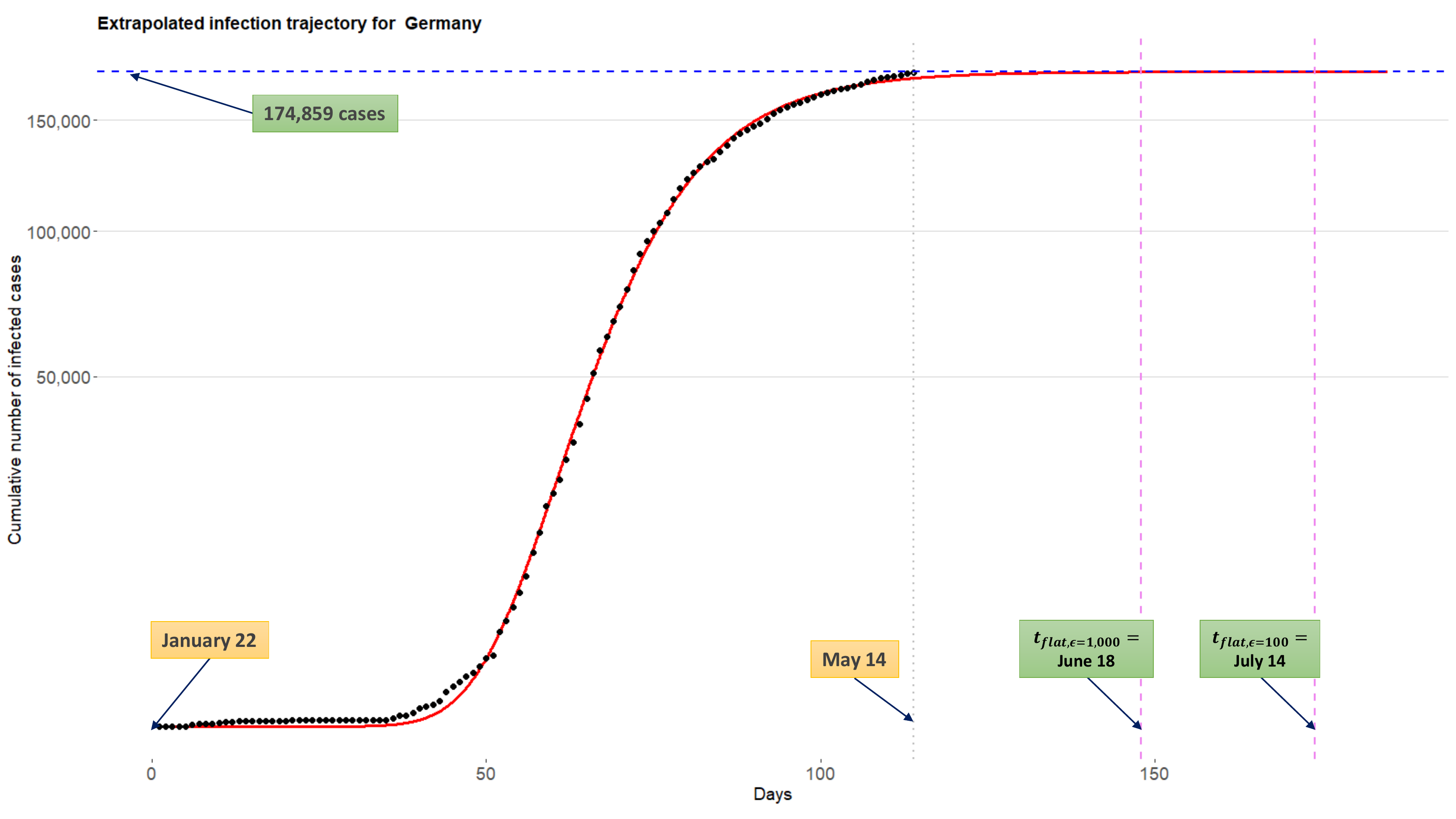}
    \caption{\baselineskip=10pt   Extrapolated infection trajectory for the Germany based on the model $\mathcal{M}_3$.}
    \label{fig:France}
\end{figure}

\begin{figure}[H]
    \centering
    \includegraphics[scale = 0.42]{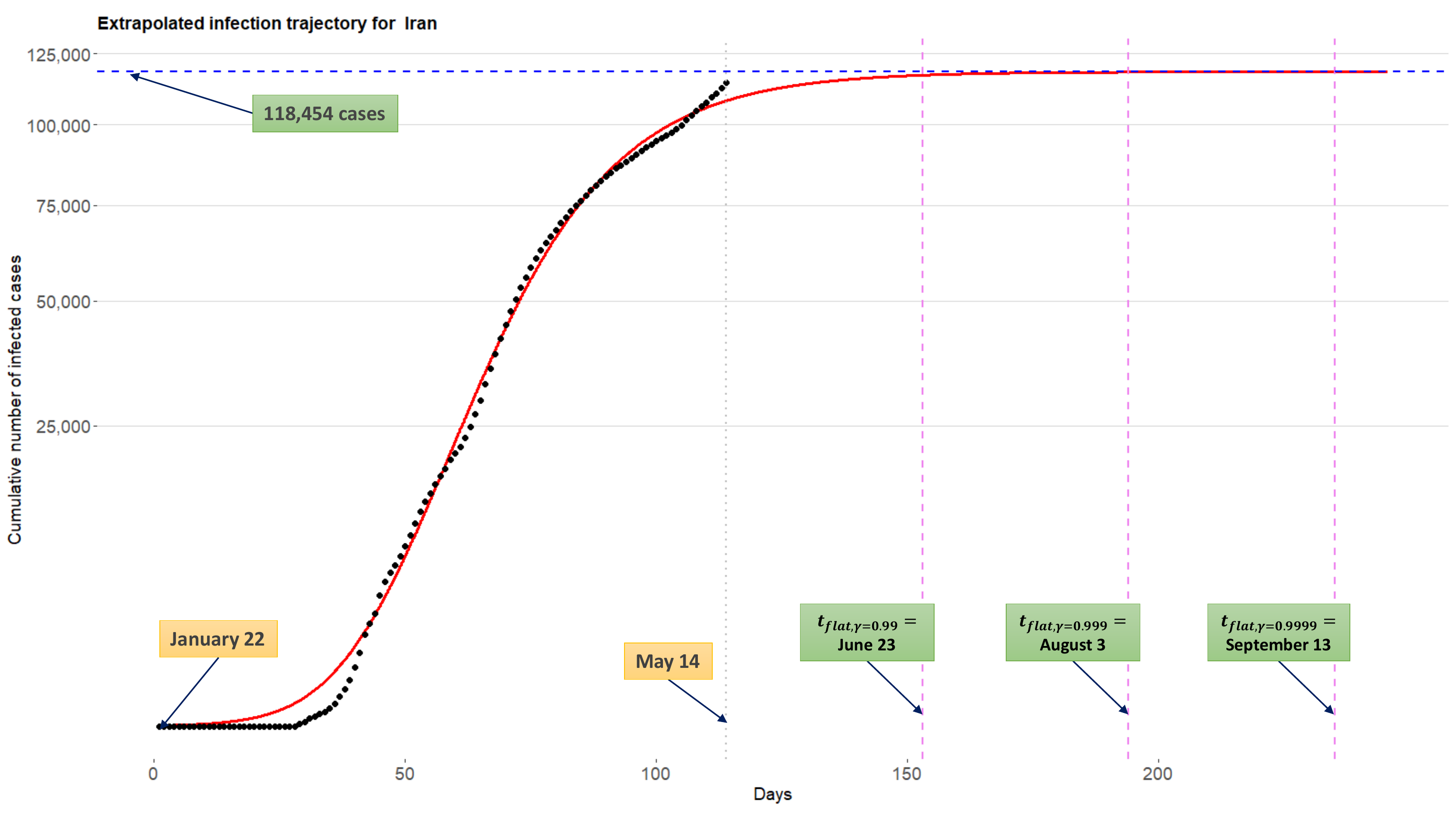}
    \caption{\baselineskip=10pt   Extrapolated infection trajectory for the Iran based on the model $\mathcal{M}_3$.}
    \label{fig:Iran}
\end{figure}

\begin{figure}[H]
    \centering
    \includegraphics[scale = 0.42]{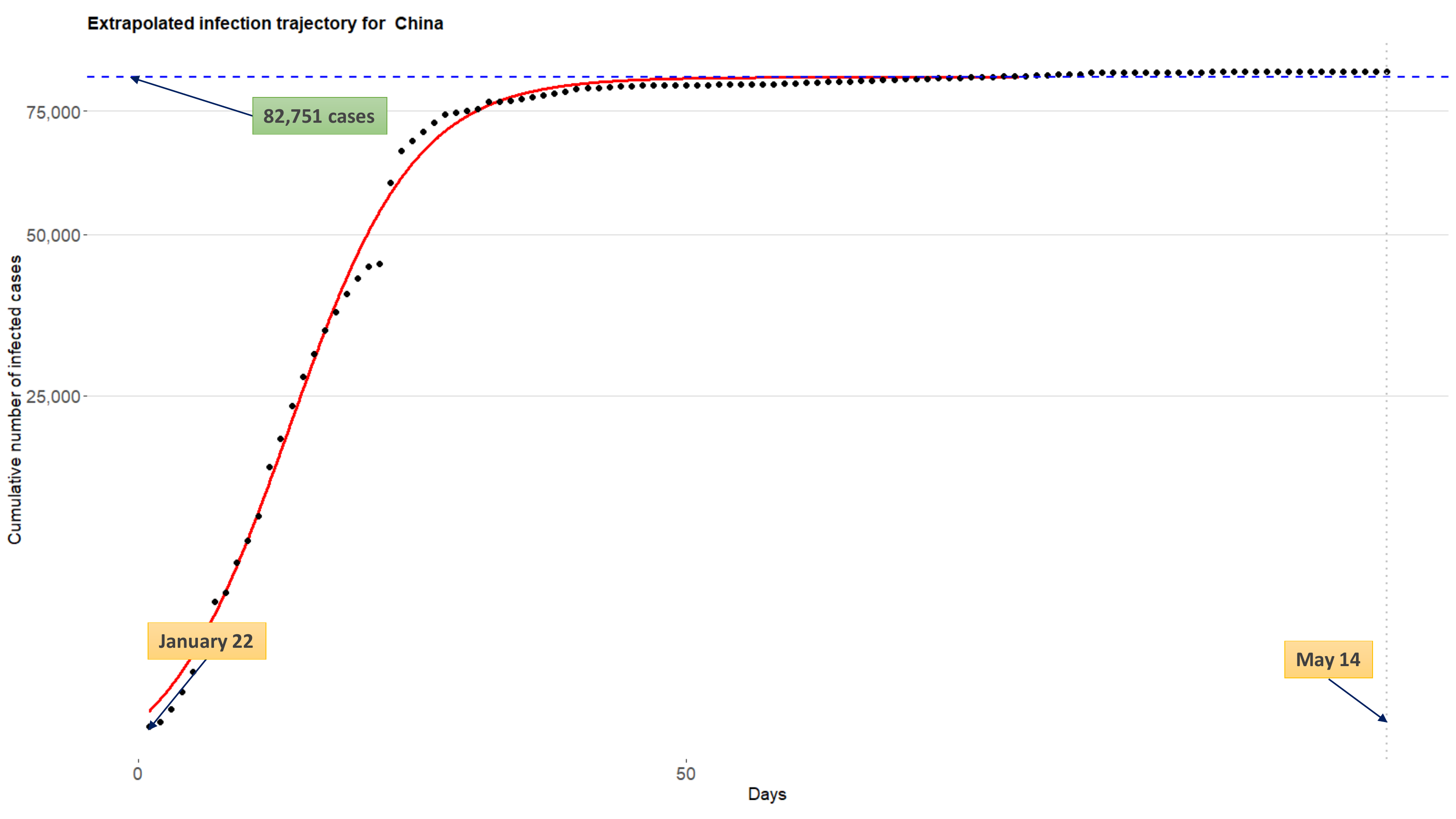}
    \caption{\baselineskip=10pt   Extrapolated infection trajectory for the China based on the model $\mathcal{M}_3$.}
    \label{fig:China}
\end{figure}

\begin{figure}[H]
    \centering
    \includegraphics[scale = 0.42]{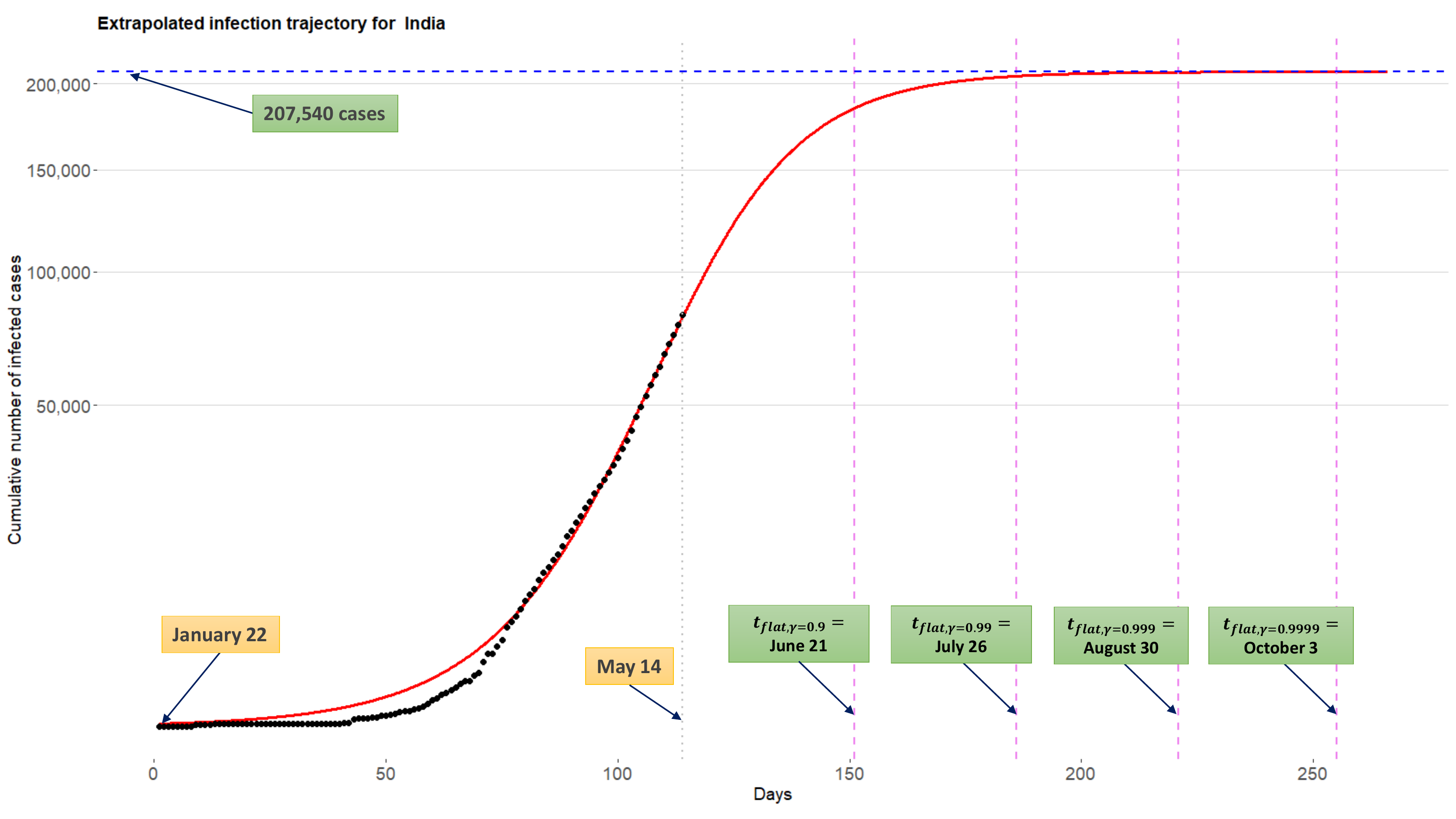}
    \caption{\baselineskip=10pt   Extrapolated infection trajectory for the India based on the model $\mathcal{M}_3$.}
    \label{fig:India}
\end{figure}

\begin{figure}[H]
    \centering
    \includegraphics[scale = 0.42]{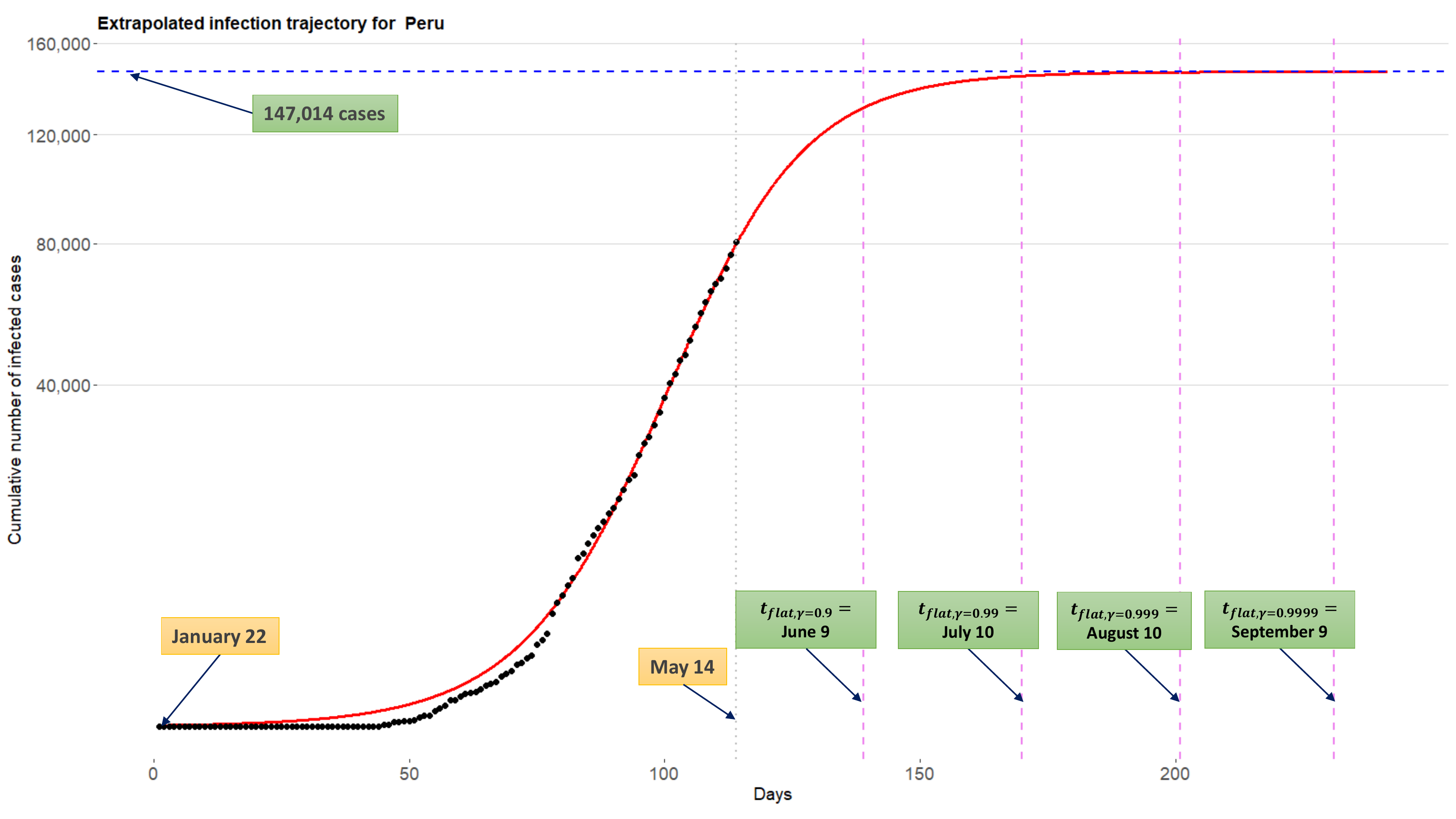}
    \caption{\baselineskip=10pt   Extrapolated infection trajectory for the Peru based on the model $\mathcal{M}_3$.}
    \label{fig:Peru}
\end{figure}

\begin{figure}[H]
    \centering
    \includegraphics[scale = 0.42]{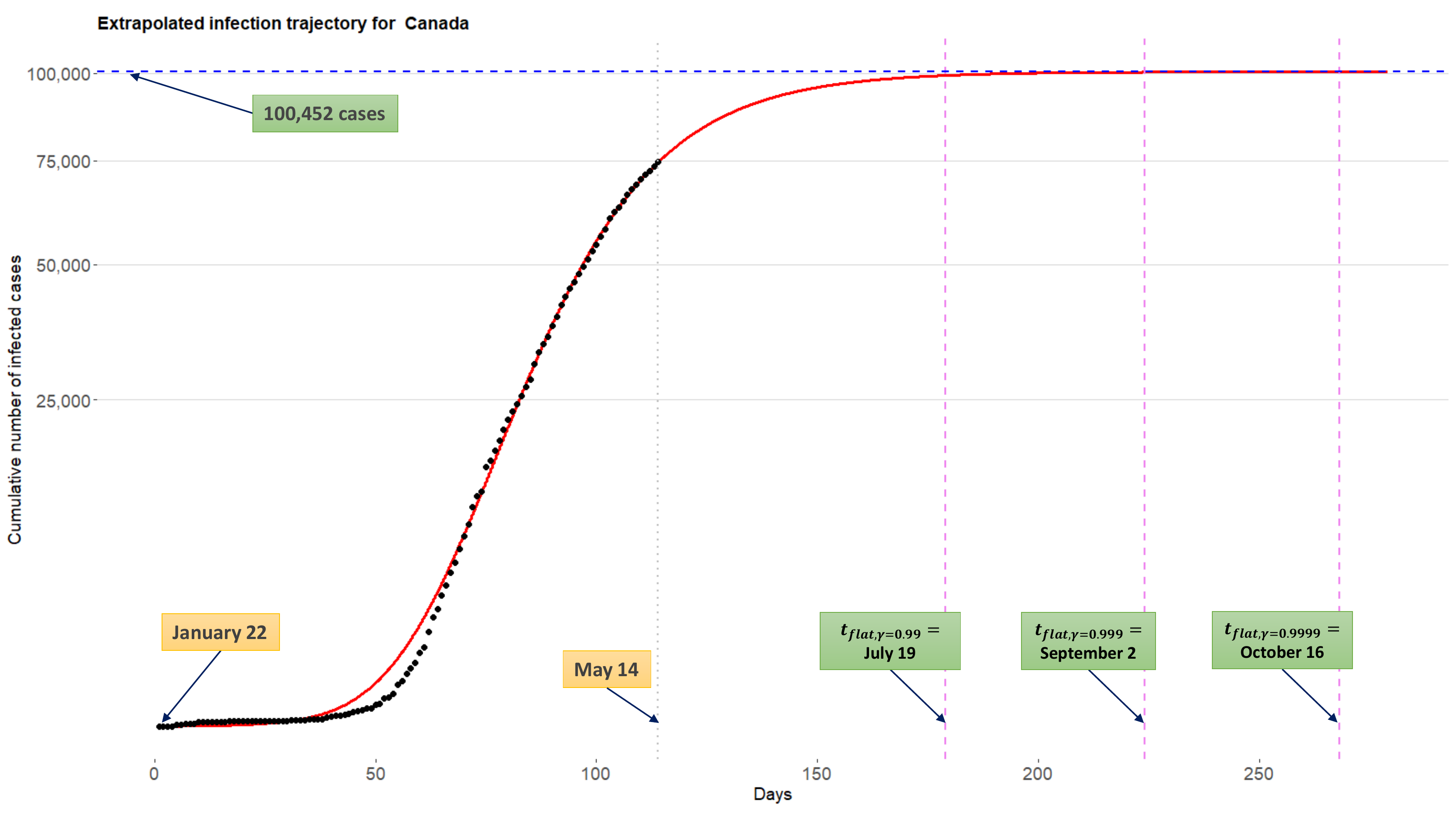}
    \caption{\baselineskip=10pt   Extrapolated infection trajectory for the Canada based on the model $\mathcal{M}_3$.}
    \label{fig:Canada}
\end{figure}

\begin{figure}[H]
    \centering
    \includegraphics[scale = 0.42]{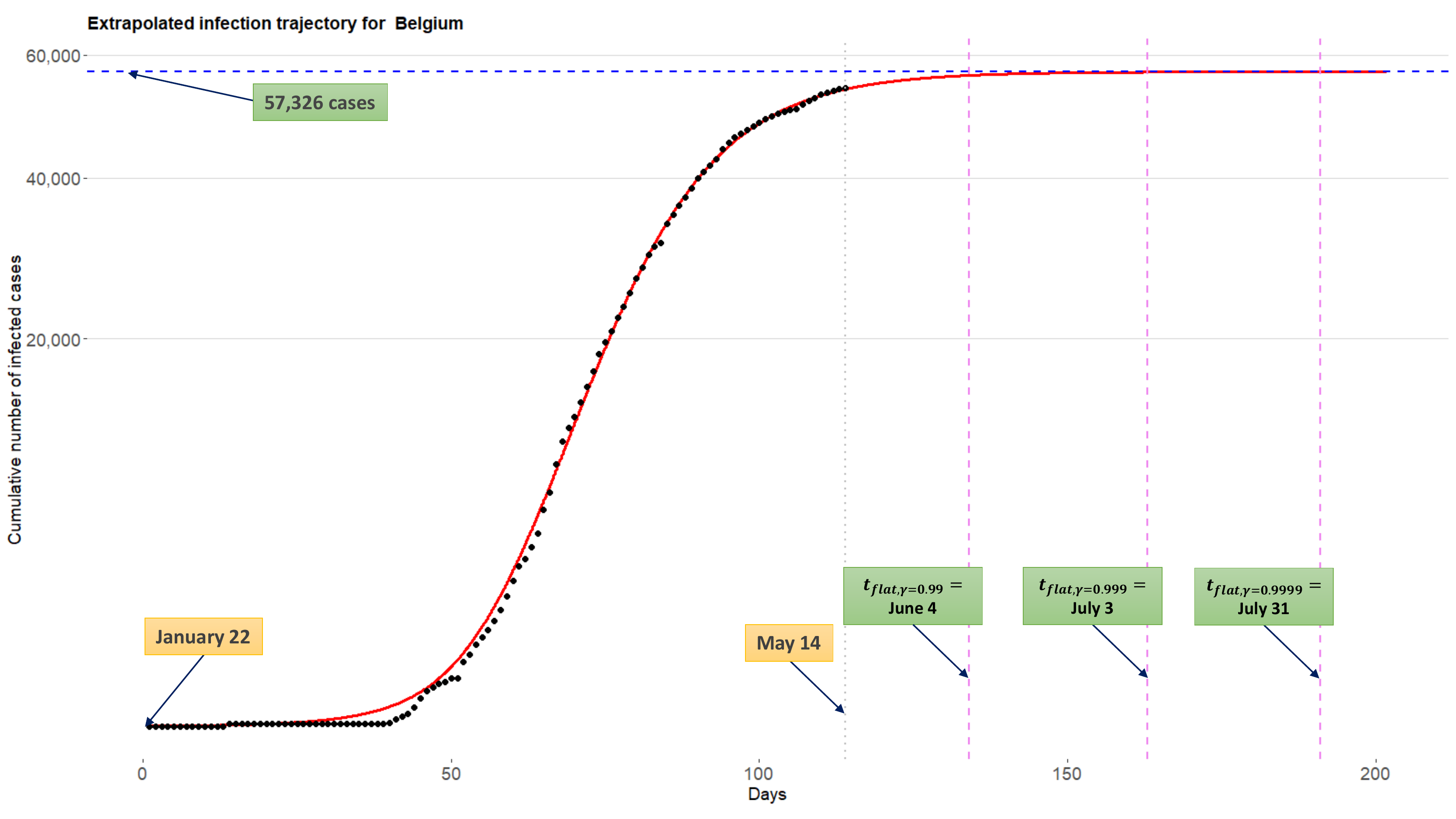}
    \caption{\baselineskip=10pt   Extrapolated infection trajectory for the Belgium based on the model $\mathcal{M}_3$.}
    \label{fig:Belgium}
\end{figure}

\begin{figure}[H]
    \centering
    \includegraphics[scale = 0.42]{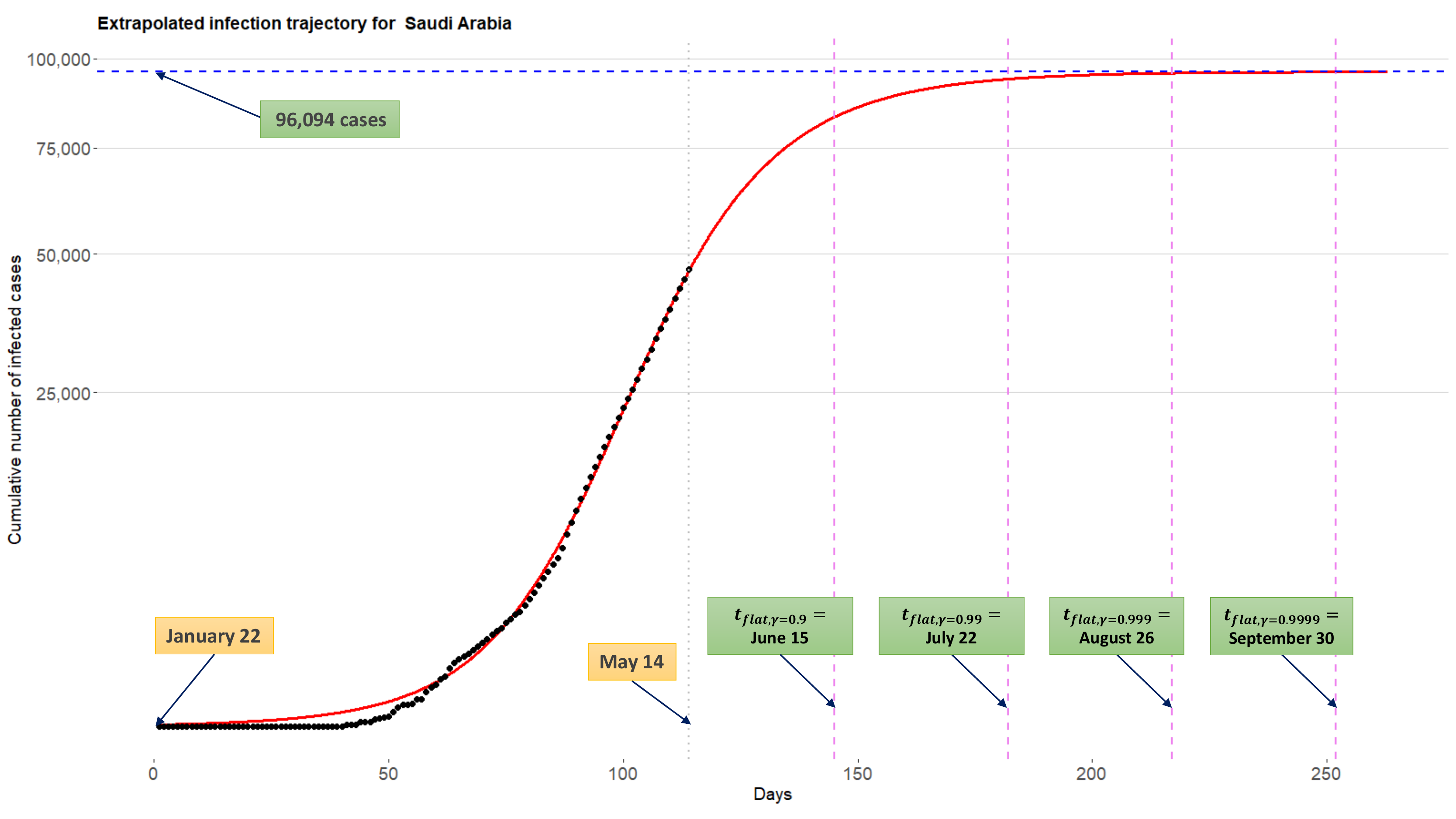}
    \caption{\baselineskip=10pt   Extrapolated infection trajectory for the Saudi Arabia based on the model $\mathcal{M}_3$.}
    \label{fig:Saudi_Arabia}
\end{figure}

\begin{figure}[H]
    \centering
    \includegraphics[scale = 0.42]{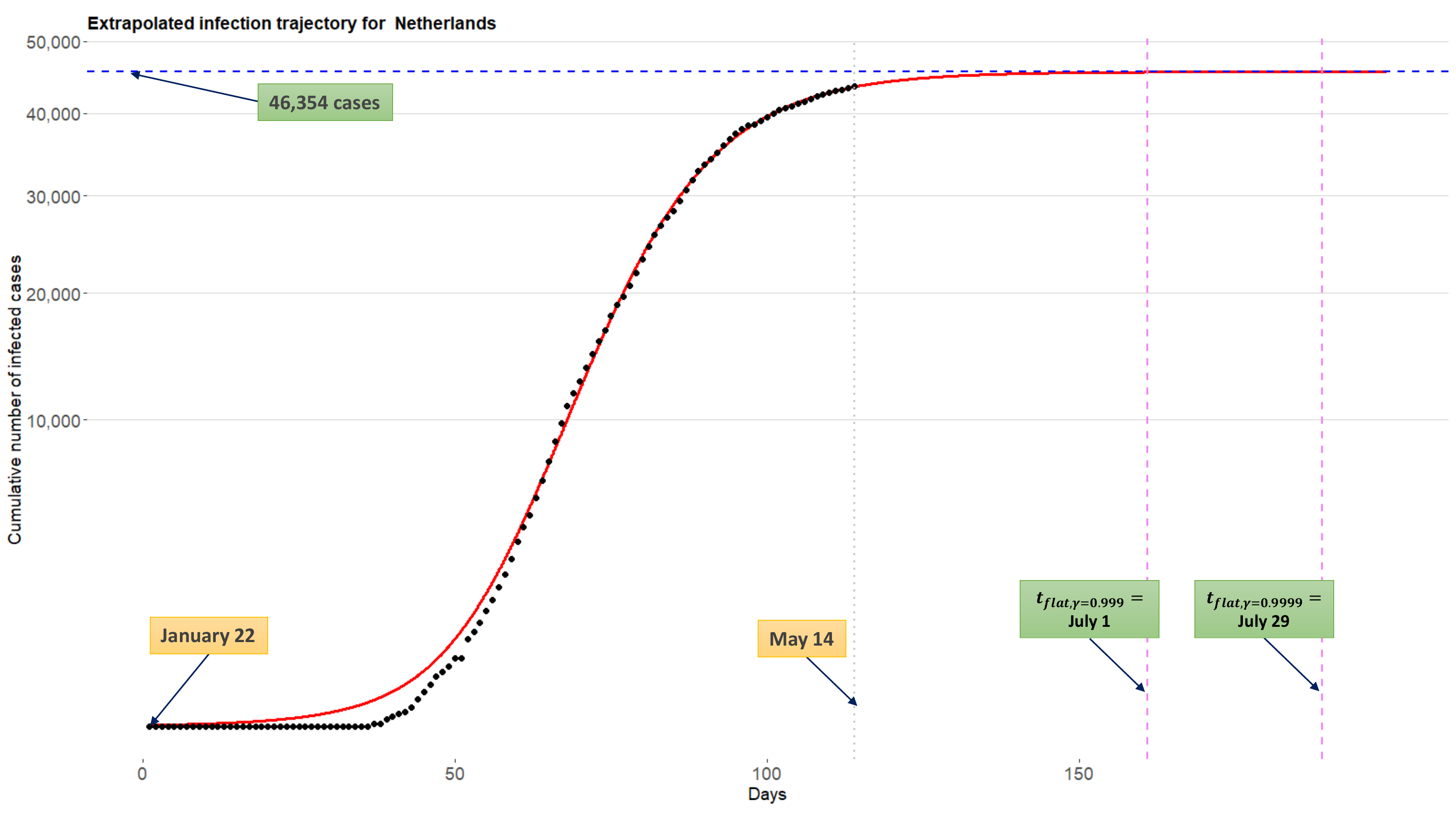}
    \caption{\baselineskip=10pt   Extrapolated infection trajectory for the Netherlands based on the model $\mathcal{M}_3$.}
    \label{fig:Netherlands}
\end{figure}

\begin{figure}[H]
    \centering
    \includegraphics[scale = 0.42]{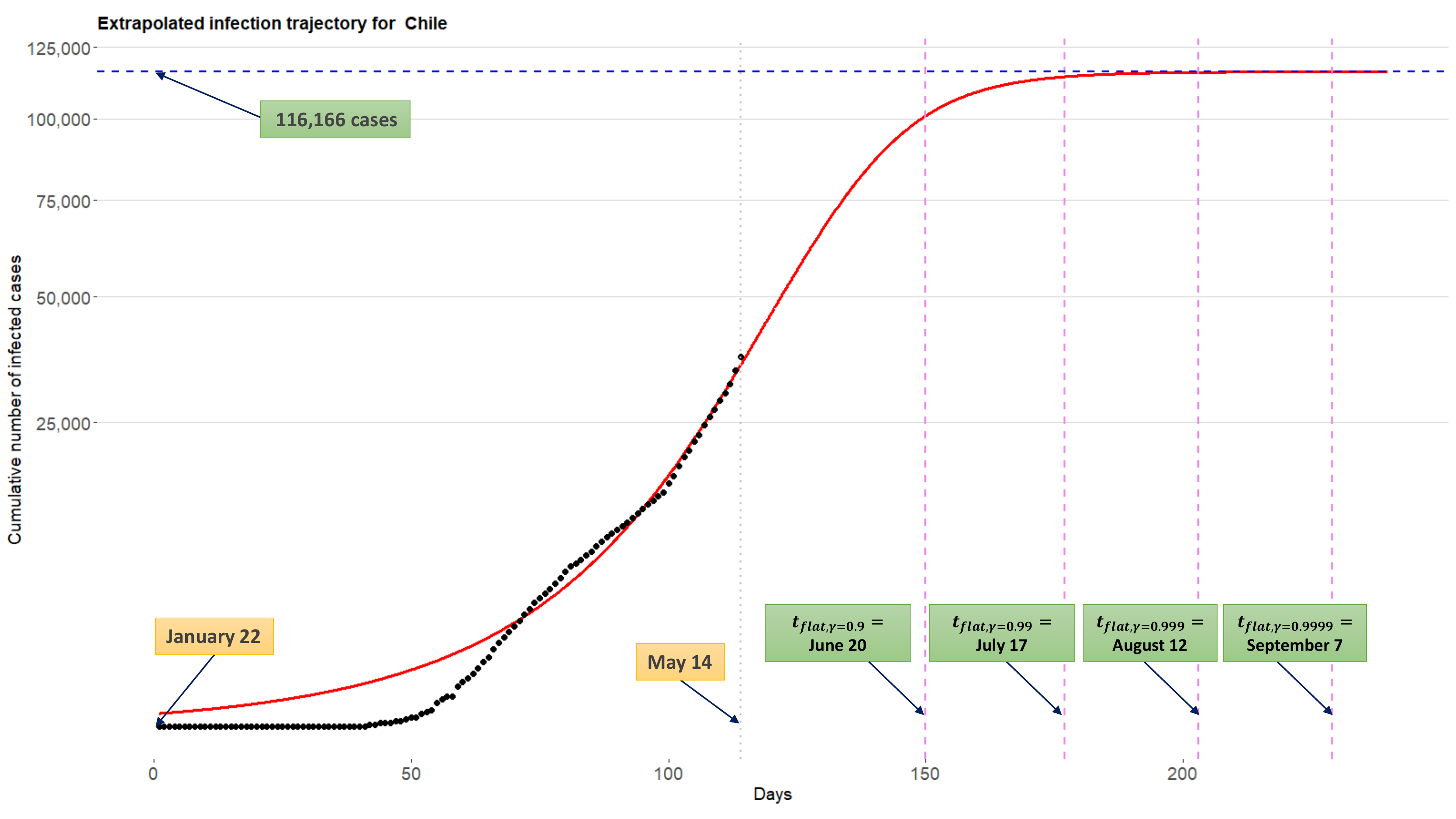}
    \caption{\baselineskip=10pt   Extrapolated infection trajectory for the Chile based on the model $\mathcal{M}_3$.}
    \label{fig:Chile}
\end{figure}

\begin{figure}[H]
    \centering
    \includegraphics[scale = 0.42]{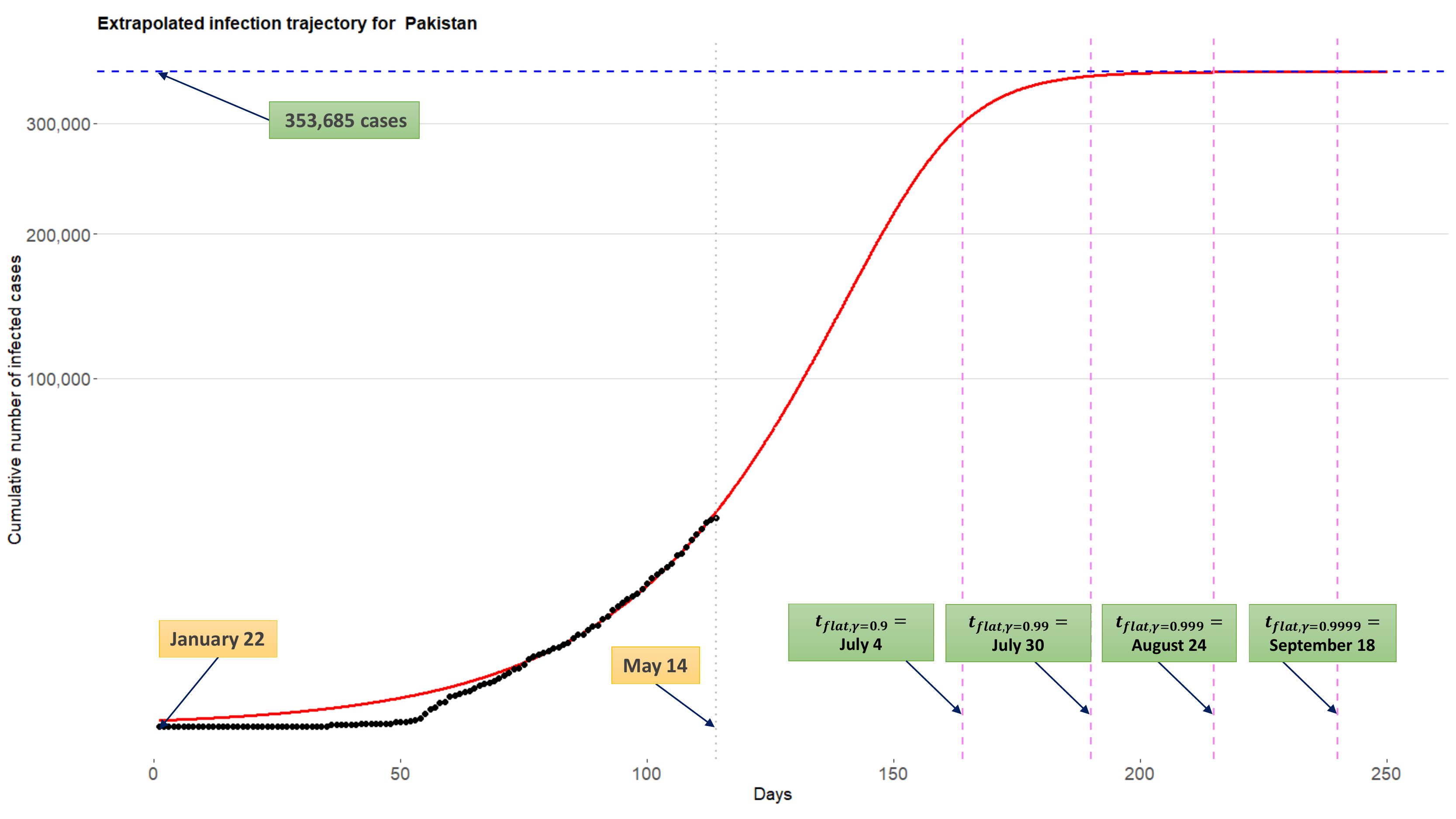}
    \caption{\baselineskip=10pt   Extrapolated infection trajectory for the Pakistan based on the model $\mathcal{M}_3$.}
    \label{fig:Pakistan}
\end{figure}

\begin{figure}[H]
    \centering
    \includegraphics[scale = 0.42]{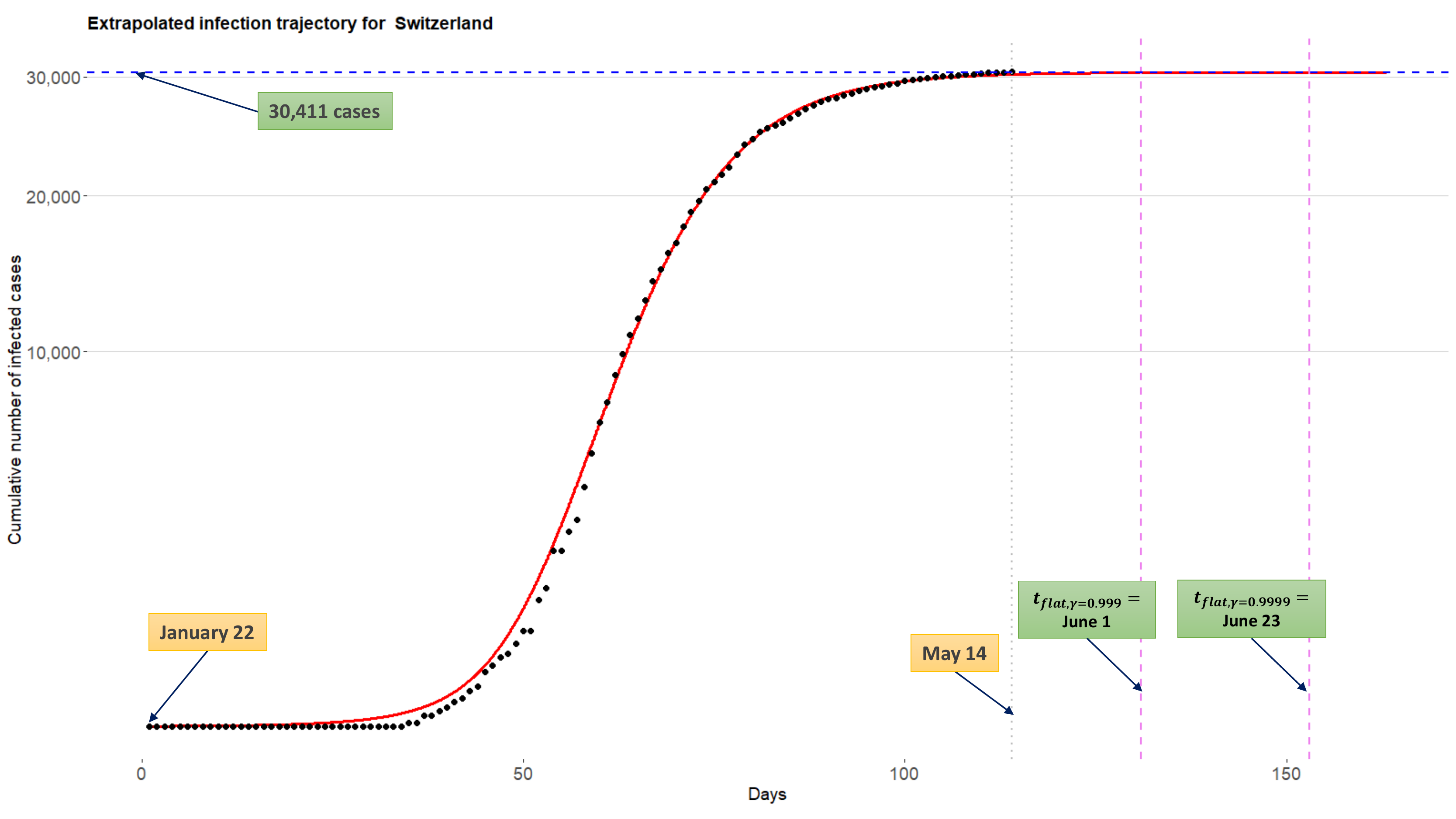}
    \caption{\baselineskip=10pt   Extrapolated infection trajectory for the Switzerland based on the model $\mathcal{M}_3$.}
    \label{fig:Switzerland}
\end{figure}

\begin{figure}[H]
    \centering
    \includegraphics[scale = 0.42]{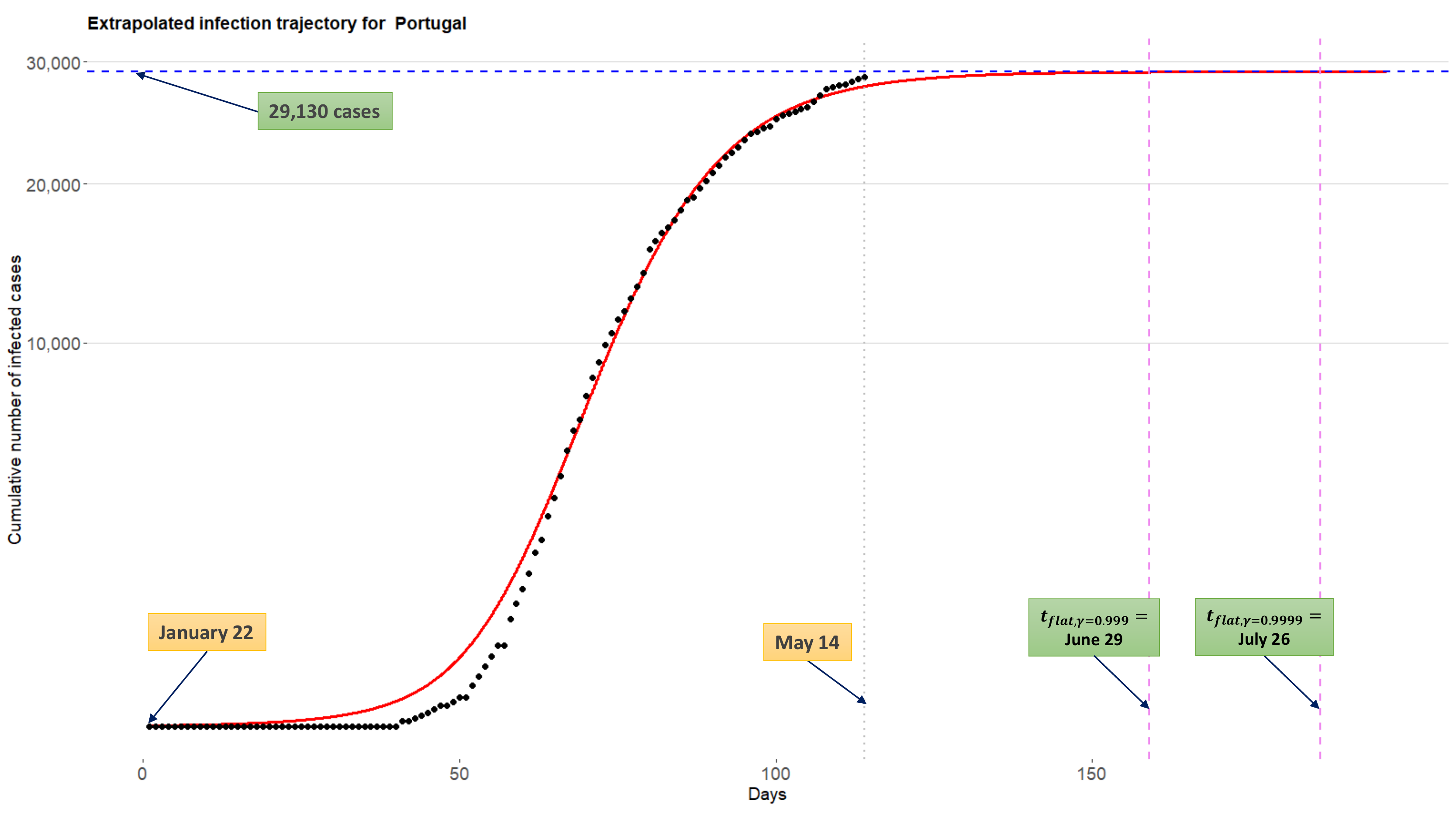}
    \caption{\baselineskip=10pt   Extrapolated infection trajectory for the Portugal based on the model $\mathcal{M}_3$.}
    \label{fig:Portugal}
\end{figure}

%\section*{Acknowledgments}

%\nolinenumbers

% Either type in your references using
% \begin{thebibliography}{}
% \bibitem{}
% Text
% \end{thebibliography}
%
% or
%
% Compile your BiBTeX database using our plos2015.bst
% style file and paste the contents of your .bbl file
% here. See http://journals.plos.org/plosone/s/latex for 
% step-by-step instructions.
% 

\end{document}